%% file: EMSE23RPython.tex
\RequirePackage{fix-cm}
\documentclass[smallextended]{svjour3}       
\smartqed  
\usepackage{graphicx}
\usepackage[authoryear]{natbib}
\AtBeginDocument{%
	\providecommand\BibTeX{{%
			\normalfont B\kern-0.5em{\scshape i\kern-0.25em b}\kern-0.8em\TeX}}}

\include{macro}

\begin{document}


\title{Characterizing Bugs in Python and R Data Analytics Programs}


\author{
	Shibbir Ahmed  \and 
	Mohammad Wardat \and 
	Hamid Bagheri \and 
    Breno Dantas Cruz \and 
	Hridesh Rajan
}


\institute{Shibbir Ahmed  \at
           Dept. of Computer Science, Iowa State University \\
           \email{shibbir@iastate.edu} 
           \and
           Mohammad Wardat  \at
           Dept. of Computer Science, Iowa State University \\
           \email{wardat@iastate.edu} 
           \and
           Hamid Bagheri \at
           Dept. of Computer Science, Iowa State University \\
           \email{bagherihamid@johndeere.com} 
           \and
           Breno Dantas Cruz  \at
           Dept. of Computer Science, Iowa State University \\
           \email{bdantasc@iastate.edu@iastate.edu} 
           \and
           Hridesh Rajan  \at
           Dept. of Computer Science, University of Central Florida \\
           \email{hridesh@iastate.edu} 
}

\date{Received: date / Accepted: date}

\maketitle

\begin{abstract}
\input{abstract}
\end{abstract}

\input{introduction.tex}

\input{methodology.tex}

\input{rq1.tex}

\input{rq2.tex}

\input{rq3.tex}

\input{rq4.tex}

\input{rq7.tex}
\input{lessons.tex}
\input{tv.tex}
\input{related.tex}
\input{conclusion.tex}
\input{data-availability.tex}

\balance

\section*{Conflict of interest}

The authors declare conflict of interest with the people affiliated with Iowa State University.

\bibliographystyle{spbasic}      
\bibliography{ICSE23/EMSE23RPython}

\end{document}

%% file: macro.tex
\usepackage{multirow}
\usepackage{caption} 
\usepackage{booktabs} 
\usepackage{graphicx}
\usepackage[belowskip=-12pt]{caption}
\usepackage{rotating}
\usepackage{tabularx}
\usepackage{subfig}
\usepackage{wrapfig}
\usepackage{listings}
\usepackage{color, colortbl}
\usepackage{balance} 
\usepackage{lscape}
\usepackage{xurl}
\usepackage{hyperref}
\usepackage{xspace}
\usepackage{framed}
\usepackage{syntax}
\usepackage[framemethod=TikZ]{mdframed}
\usepackage{soul}
\usepackage{flushend}
\usepackage{subfig}
\definecolor{codegreen}{rgb}{0,0.6,0}
\definecolor{codegray}{rgb}{0.5,0.5,0.5}
\definecolor{codepurple}{rgb}{0.58,0,0.82}
\definecolor{backcolour}{rgb}{0.95,0.95,0.92}
\definecolor{Gray}{gray}{0.1}
\lstdefinestyle{mystyle}{
	backgroundcolor=\color{backcolour},   
	commentstyle=\color{codegreen},
	keywordstyle=\color{magenta},
	numberstyle=\tiny\color{codegray},
	stringstyle=\color{codepurple},
	basicstyle=\scriptsize,
	breakatwhitespace=false,         
	breaklines=true,                 
	captionpos=b,                    
	keepspaces=true,                 
	numbers=left,                    
	numbersep=5pt,                  
	showspaces=false,                
	showstringspaces=false,
	showtabs=false,                  
	tabsize=2
}

\lstdefinelanguage{Pythonna}{%
	language     = python,
	morekeywords = {to_categorical, flow_from_directory,pad_sequences,load_image}
}

\lstdefinestyle{customc}{
	belowcaptionskip=1\baselineskip,
	breaklines=false,
	frame= single,
	breaklines = true,
	xleftmargin=\parindent,
	language= Pythonna,
	showstringspaces=false,
	basicstyle=\footnotesize\ttfamily,
	keywordstyle=\bfseries\color{green!40!black},
	commentstyle=\itshape\color{purple!40!black},
	identifierstyle=\color{blue},
	stringstyle=\color{codegreen},
	backgroundcolor=\color{gray!4},
}
\usepackage{hyperref}
\usepackage{xcolor}
\definecolor{medium-blue}{rgb}{0,0,1}
\hypersetup{colorlinks, urlcolor={medium-blue}}

\graphicspath{{./figures/}}
\newcommand{\figref}[1]{Fig.~\ref{#1}}
\newcommand{\tabref}[1]{Table~\ref{#1}}

\newcommand{\etal}{{\em et al.}\xspace}

\newcommand{\sof}{\textit{Stack Overflow}\xspace}

\newcommand{\ggplot}{\textit{ggplot2}\xspace}
\newcommand{\dplyr}{\textit{dplyr}\xspace}
\newcommand{\shiny}{\textit{shiny}\xspace}
\newcommand{\datatable}{\textit{data.table}\xspace}

\newcommand{\matrixr}{\textit{matrix}\xspace}
\newcommand{\timeseries}{\textit{time-series}\xspace}
\newcommand{\plotly}{\textit{plotly}\xspace}
\newcommand{\tidyr}{\textit{tidyr}\xspace}

\newcommand{\igraph}{\textit{igraph}\xspace}
\newcommand{\rcaret}{\textit{r-caret}\xspace}					
\newcommand{\randomforest}{\textit{randomforest}\xspace}								
\newcommand{\CRAN}{\textit{CRAN}\xspace}					
\newcommand{\BioConductor}{\textit{BioConductor}\xspace}					
\newcommand{\RForge}{\textit{R-Forge}\xspace}

\newcommand{\numpy}{\textit{Numpy}\xspace}
\newcommand{\pandas}{\textit{Pandas}\xspace}
\newcommand{\dataframe}{\textit{Dataframe}\xspace}
\newcommand{\matplotlib}{\textit{Matplotlib}\xspace}

\newcommand{\scikilearn}{\textit{Sciki-Learn}\xspace}
\newcommand{\scipy}{\textit{SciPy}\xspace}

\newcommand{\pyqt}{\textit{PyQt}\xspace}

\newcommand{\tensor}{\textit{Tensorflow}\xspace}

\newcommand{\gh}{\textit{GitHub}\xspace}

\newcounter{rqs}
\stepcounter{rqs}

\newcounter{NumObservations}
\stepcounter{NumObservations}
\definecolor{shadecolor}{rgb}{.9,.9,.9}

\newcommand{\finding}[1]{%
	\begin{mdframed}[
		backgroundcolor= cyan!05,
		linewidth=1pt,
		linecolor = black!50,
		roundcorner=5pt,
		innertopmargin=2pt,
		innerbottommargin=0pt,
		innerrightmargin=4pt,
		innerleftmargin=2pt,
		leftmargin = 0pt,
		rightmargin = 0pt]%
		\noindent{\textbf{Finding \arabic{NumObservations}}: \em #1}
	\end{mdframed}%
	\stepcounter{NumObservations}
}

\definecolor{diffstart}{rgb}{1,1,1}
\definecolor{diffincl}{rgb}{0,.42,.24}
\definecolor{diffrem}{rgb}{.8,0,0}

\usepackage{listings}
  \lstdefinelanguage{diff}{
    basicstyle=\fontsize{6}{6}\ttfamily,
    morecomment=[f][\color{diffstart}]{@@},
    morecomment=[f][\color{diffincl}]{+\ },
    morecomment=[f][\color{diffrem}]{-\ },
  }

%% file: abstract.tex
R and Python are among the most popular languages used in many critical data analytics tasks. However, we still do not fully understand the capabilities of these two languages regarding bugs
encountered in data analytics programs. 
What type of bugs are common? What are the main root causes? 
What is the relation between bugs and root causes? How to mitigate these bugs?
We present a comprehensive study of {\bf 5,068} \sof posts, 
{\bf 1,800} bug fix commits from \gh repositories, and several \gh issues of the most used libraries to understand bugs in R and Python.
Our key findings include: 
while both R and Python programs have bugs due to inexperience with data analysis, 
Python see significantly larger data preprocessing bugs compared to R.
Developers experience significantly more data flow bugs in R 
because intermediate results are often implicit. 
We also found changes in packages and libraries cause more bugs in R 
compared to Python while package or library mis-selection and conflicts 
cause more bugs in Python than R.
In terms of data visualization, R packages have significantly more bugs than Python libraries.
We also identified  a strong correlation between comparable packages in R 
and Python despite their linguistic and methodological differences.
Lastly, we contribute a large dataset of manually verified R and Python bugs.

%% file: introduction.tex
\section{Introduction}
\label{sec:introduction}
Data Analytics programs~\cite{runkler2020data} are being used for several tasks, such as data acquisition, manipulation, visualization, mathematical computations, and machine learning~\cite{mitchell1999machine}. These programs find relevant information, structures, and patterns, to obtain insights, identify causes and effects, make predictions, or suggest optimal decisions~\cite{runkler2020data}.
Due to their utilization in a wide range of domains (i.e., medical, financial, scientific, etc.), data analytics bugs can have severe consequences such as misdiagnoses~\cite{nuemdMedical,khnMedical}, stock market misinterpretations~\cite{dataversity}, bias decisions~\cite{yahooFairness}, incorrect scientific results~\cite{bbcAI}. 
Unlike in traditional programs, data analysts face the challenge of data quality~\cite{8046093}; therefore, debugging those software requires reasoning not only about the code but also about the data. However, the new bug types in data analytics software remain not fully understood by the community. 
R~\cite{R} and Python~\cite{sanner1999python} are among the most popular programming languages for building software by data analysts. To the best of our knowledge, there has not been a comprehensive study on bugs in data analytics programs. To address this issue, this work performs a comprehensive study of bugs in data analytics tasks. Even though R and Python have different programming paradigms, we target these languages because they are the most popular for those tasks, with developers using them to solve similar problems. 
In particular, we studied posts that utilized the eleven most tagged data analytics packages for R on \sof: \ggplot, \dplyr, \shiny, \datatable, \matrixr, \timeseries, \plotly, \tidyr,\rcaret, \randomforest, and \igraph. Also, we studied posts that used the seven most tagged data analytics packages for Python on \sof: \matplotlib, \pyqt, \dataframe, \pandas, \numpy, \scipy, and \scikilearn. These packages are the most popular to address these functionalities and have the most downloads. Also, those packages are most frequently tagged libraries on \sof. We studied issues on \gh repositories of these libraries over eight years.
The datasets used in this study include a significantly large number of high-quality \sof posts and bug fix commits in open source repositories on \gh. We manually studied 1,299 \sof posts about the collected R packages and 1,100 bug-fixing commits in R code from \gh. In total, we found 559 and 544 bugs, respectively. Similarly, we studied 3,769 \sof posts and 700 \gh fix commits from the collected Python packages. We observed that 1,210 and 1,132 were bug-related, respectively. We have manually studied a total of 5,068 \sof posts and 1,800 \gh fix commits. For the \sof dataset, we did not take a sample. The methodology section describes the post-selection criteria. For only the \gh dataset, we took 100 random samples from all packages. We have analyzed this data to answer the following research questions:

\begin{itemize}
\item[RQ1]\textbf{(Classes of Bugs)}: What classes of bugs are the most frequent in data analytics programs? 
\item[RQ2]\textbf{(Frequent Root Causes)}: What are the common root causes of bugs in data analytics programs? 
\item[RQ3]\textbf{(Relationship among bug
types and root causes and effects)}: What is the relationship among bugs and root causes and effects in different data analytics tasks?
\item[RQ4]\textbf{(Effect)}: What are the persistent effects of bugs in data analytics programs?
\item[RQ5] \textbf{(Commonality of Bugs)} Do different data analytics libraries have commonalities in terms of types of bugs?
\end{itemize}

We have found that Logic Bugs (LB) are the most frequent bugs in R and Python, Data Bugs (DB) appear significantly more in Python while Data Flow Bugs (DFB) appear more often in R than in Python. Regarding root causes of bugs in data analytics software, we have found that  Confusion with Data Analysis (CDA) is the most common root cause of bugs in both languages. Incorrect Data Analysis Parameters (IDAP) is the second most common root cause of bugs. We found changes and bugs in packages (APIC) cause more bugs in R than in Python. Package/Library Mis-selection and Conflicts (PLMC) cause more bugs in Python than R. While R has a slightly higher readability barrier for data analysts, the statistical power of R led to less number of bad performance bugs. In Python, the new version of libraries caused bad performance. Also, we have found a strong correlation between comparable packages in R and Python for similar data analytics tasks despite several differences.

%% file: methodology.tex
\section{Methodology}
\label{sec:methodology}
In this section, we present the data collection methodology for \sof and \gh. We provide criteria for selecting packages and bug-fixing commits. We describe our proposed taxonomy to label bugs, causes, and effects for R and Python data analytics programs. Lastly, we discuss the dataset labeling methodology.


\subsection{Data Collection}
To select the dataset for our study, first, we select posts discussing data analytics packages in R and Python. Then, we collect the frequencies of tags of R and Python packages in posts of \sof, starting in 2008. Then, we remove non-package tags, while verifying if they are among the top downloaded packages on \CRAN. Lastly, we chose eleven packages: \ggplot~\cite{ggplot2}, \dplyr~\cite{dplyr}, \shiny~\cite{shiny}, \datatable~\cite{data.table}, \matrixr~\cite{shiny}, \timeseries~\cite{shiny}, \plotly~\cite{plotly}, \tidyr~\cite{tidyr},\rcaret~\cite{caret}, \randomforest~\cite{randomForest}, and \igraph~\cite{igraph}. The selection ensures that those are the eleven most tagged data analytics packages in R on \sof since 2008. By applying similar methodology for Python, we selected \matplotlib \cite{matplotlib}, \pyqt~\cite{pyqt}, \dataframe~\cite{dataframe}, \pandas~\cite{pandas}, \numpy~\cite{numpy}, \scipy~\cite{scipy}, and \scikilearn~\cite{scikit} that are the top seven packages by posts frequencies in \sof.  We chose 11 R and 7 Python packages because we wanted data points for similar functionality, e.g., for visualization tasks 739 and 771 posts in R and Python. Therefore, if we took the top seven packages for each language, then we would have had unbalanced datasets. Since Python is a general-purpose language, we removed the libraries and frameworks that do not have data analytics-related functionalities (i.e., Django and flask). Also, we have not selected machine learning-specific libraries such as \tensor because there has been recent work studying them~\cite{islam2019comprehensive}. 
\input{tables/SOGH}
\tabref{tbl:dataset} shows a summary of the datasets. In particular, we have studied two datasets of 5,068 \sof posts that discuss bugs and 1,800 fix commits in the \gh repositories that are using the top used packages in R and Python. We randomly selected 100 fix commits for each library to restrict the manual efforts and perform a quality study.

\subsubsection{\sof Data Collection}
We downloaded all the \sof posts with score $\geq 10$ that have code snippets for the top R and Python packages because when developers face a bug, chances are that they post the code snippet on the \sof to get help from others.  
\sof provides an API to retrieve posts that contain codes. We applied a filter to the posts with scores that are $\geq 10$ to select the high-quality posts. We added an additional filter by checking the body of the posts if they contain the words \texttt{"bug"}, \texttt{"error"}, \texttt{"fail"}, \texttt{"not work"}, \texttt{"performance"}, \texttt{"expect"}, \texttt{"crash"}, \texttt{"incorrect"}, and \texttt{"fault"}. We borrowed this idea from the literature~\cite{beizer1984software,zhang2018empirical,islam2019comprehensive} that used similar keywords in filtering \sof and \gh commits. After downloading all \sof posts, we manually checked all the posts to make sure posts are discussing bugs, and they are not new features or general questions about R and Python. This avoids the problem of bug misclassification as stated by the previous work~\cite{10.5555/2486788.2486840}.


\subsubsection{\gh Data Collection}
\label{subsec:gh-data}
To select the \gh repositories for our study, we followed similar selection criteria from prior work~\cite{islam2019comprehensive, islam20repairing}. In particular, we selected repositories from the last 12 years, which include the studied 11 R and 7 Python packages containing the ``fix'' keyword in the commit history.
We checked manually that the code imported the data analytics libraries. This may result in false positives. To remedy this problem, we manually checked the code to make sure it has the correct library imported into the code. The bug-fixing commits involved both direct and indirect libraries. Then, we manually looked over all commits line by line to find bug fixes. In several cases, one commit contained more than one bug fix. In some commits, developers provided a link to the issue in the title or commit messages, and we read the issues to make sure the commit is fixing a bug, not just adding new functionality or improving their style of code. 
The margin of error for randomly selecting GitHub python and R projects is 95\%.
We studied 1903 closed GitHub issues over the last 8 years to determine the deprecation history of Python and R visualization packages in response to resolved GitHub issues on API change, package update, and version issues. We studied 1068 closed Git issues to analyze the deprecation history of Python and R packages.
\subsubsection{Balanced dataset test}
We used Shannon Entropy~\cite{hong2016dealing}, which is a value between 0 to 1, to measure whether our dataset, i.e., bugs in Table~\ref{tbl:dataset}, is balanced or imbalanced. An imbalanced dataset means that the numbers of observations for different classes are significantly different. The values of entropy show that both R and Python have large numbers around 0.77; therefore, our R and Python datasets do not have an imbalanced challenge.
\subsection{Taxonomy of Bugs, Root Causes, and Effect}
\label{subsec:classification}
We give three labels, i.e., bug types, root cause, and effect, to each \sof post and \gh fix commit. We used the well-vetted taxonomy of bugs from prior works~\cite{beizer1984software, zhang2018empirical, islam2019comprehensive} and extended it during the pilot study while following open card sorting methodology~\cite{fincher2005making}. The final classification includes important aspects of data analytics such as Incorrect Data Analysis Parameters, Confusion with Data Analysis, and Data Analysis Inefficiency as root causes. For bug types, we expand the meaning of Logic Bugs to include confusion in stat/math, domain knowledge, and steps of data analysis pipelines. 
Also, we have categorized packages in R and Python based on similar tasks. For example, in R, \ggplot, \plotly, \shiny, and \igraph are used for data visualization tasks; similarly, \matplotlib and \pyqt in Python, are used by data analysts. This taxonomy of bug types, causes, and effects can help R and Python practitioners improve their understanding of data analytics programs.
We performed a pilot study as the initial step in our labeling process to determine if we need additional classes for the domain of data analytics programs that are not in the literature~\cite{beizer1984software, zhang2018empirical, islam2019comprehensive, islam20repairing}.

\textbf{Classification of Bug Types:} 
\tabref{tab:bugtypes} shows the bug classification. To provide a general classification, we studied \sof posts for data analytics programs in R and Python. Usually, posts related to bugs contain codes and error messages which include crashes, run-time errors, and bad performance. 
We come across 7 bug types except for IIS bugs from prior work~\cite{zhang2018empirical, islam2019comprehensive} because we encountered IIS-related bugs as a new category in data analytics program in our dataset during the pilot study.
\input{bugtypeClassification.tex}

\textbf{Classification of Root Causes of Bugs:} 
\tabref{tab:rootcauseClassificationTable} shows a classification specific for the root causes of bugs in data analytics programs. We focus on different granularity to identify the scale of cause and needed fix changes. 
In terms of root causes, we select Package/Library Mis-selection or conflicts (PLMC) because in these programs we get wrong choices of packages or libraries, conflicts between packages, and missing package installation or updating issues. We also include Shape Mismatch, Type Mismatch, Incorrect Data Analysis Parameter (IDAP), Misuse Required Parameter (MRP), Confusion with Data Analysis (CDA), and Data Analysis Inefficiency (DAI). The rationale behind adding these new categories, especially for Data Analytics programs is included in Table~\ref{tab:rootcauseClassificationTable}.
\input{rootcauseClassification.tex}
\input{effectsClassification.tex}
\textbf{Classification of Effects of Bugs:} \tabref{tab:effectClassification} contains classification of bug effects. 
We have not encountered any new class of effects in Table~\ref{tab:effectClassification}, thus we have not added any new effect for the data analytics domain of problems.

\subsection{Labeling Process}
\label{subsec:label-packages}
We used the following procedure to handle disagreements and decisions during the qualitative analysis. We used Cohen's Kappa coefficient~\cite{viera2005understanding} to calculate the inter-rater agreement in the labeling process for \gh and \sof datasets. The Kappa coefficient is widely used to test inter-rater reliability.
For each language, a team of two experts was assigned to label R and Python \sof posts and \gh commits. Two experts in each team are graduate students, co-authors, with several years of programming experience. Initially, we had a meeting for each group in the presence of an expert describing the classification schema for bugs, causes, and effects. We needed the expert to obtain feedback on the classification and follow the open card sorting methodology~\cite{fincher2005making}. We took the initial labels from the literature~\cite{beizer1984software, zhang2018empirical, islam2019comprehensive} and extended it during the pilot study while following open card sorting methodology and included data analytics aspects. In open card sorting, the classification schema changes during the labeling process. We started labeling 1\%  of the dataset, i.e., 1\% of each package, where each rater labeled \sof posts separately. For the packages with a smaller number of posts, we took one data point during the first training session. After that, we met and reconciled the differences and calculated the inter-rater agreement between the raters.  
During the first 1\% of the dataset labeling, the inter-rater agreement as measured by Cohen's Kappa coefficient was low due to the raters being in the training stage. After reconciling differences and checking inter-rater agreement, the Kappa coefficient improved. The team discussed all disagreements in the presence of a moderator and continued labeling the next 5\% of the dataset, resulting in a Kappa coefficient of 0.1 for R and 5\% for Python. Further training sessions were conducted to enhance understanding of the classification schema, and disagreements were resolved through discussion until a consensus was reached on bug type, cause, and effect. The same approach was followed for the next 10\% of the dataset, with close monitoring of the inter-rater agreement. The Kappa coefficient for bug type in 10\% of the Python dataset was 0.30, root cause 0.50, and effects 0.70. In 30\% of the dataset, bug type was 0.30, root cause 0.36, and effects 0.60. And in 40\% of the dataset, bug type was 0.31, root cause 0.46, and effects 0.68. All disagreements were resolved in the presence of a moderator. As more posts were labeled, inter-rater agreement among team members improved, reaching a near-perfect agreement with a Kappa coefficient above 0.80. The same methodology was applied to \gh bug fixes.

%% file: tables/SOGH.tex
\begin{table}[!th]
	\setlength{\parskip}{.2cm}
	\setlength{\belowcaptionskip}{.1cm}
\fontsize{8.5}{8.5}\selectfont
	\centering
	\caption{\gh and  \sof dataset ~\cite{dataset} summary}
	\setlength{\tabcolsep}{1.5pt}
	\begin{tabular}{|c|p{6.6em}|r|r|r|r|}
		\hline
		\multicolumn{1}{|c|}{\textbf{Language}} & \textbf{Library} & \multicolumn{2}{c|}{\textbf{Stack Overflow}} & \multicolumn{2}{c|}{\textbf{GitHub}} \\
			\cline{3-6} 		& \multicolumn{1}{r|}{} & \multicolumn{1}{c|}{\#Posts} & \multicolumn{1}{c|}{\#Bugs} & \multicolumn{1}{c|}{\#Commits} & \multicolumn{1}{c|}{\#Bugs} \\
		\hline    \multicolumn{1}{|c|}{\multirow{12}[2]{*}{\textbf{R}}} & 
		\textbf{data.table} & 234   & 98    & 100   & 29 \\
		\cline{2-6}          & \textbf{dplyr} & 200   & 114   & 100   & 35 \\
		\cline{2-6}          & \textbf{tidyr} & 14    & 9     & 100   & 10 \\
		\cline{2-6}          & \textbf{matrix} & 45    & 26    & 100   & 65 \\
		\cline{2-6}          & \textbf{randomforest} & 22    & 15    & 100   & 83 \\
		\cline{2-6}          & \textbf{r-caret} & 16    & 11    & 100   & 58 \\
		\cline{2-6}          & \textbf{time-series} & 29    & 16    & 100   & 57 \\
		\cline{2-6}          & \textbf{ggplot2} & 568   & 211   & 100   & 50 \\
		\cline{2-6}          & \textbf{plotly} & 22    & 12    & 100   & 83 \\
		\cline{2-6}          & \textbf{shiny} & 125   & 41    & 100   & 42 \\
		\cline{2-6}          & \textbf{igraph} & 24    & 6     & 100   & 32 \\
		\cline{2-6}          & \textbf{Total} & \textbf{1299}    & \textbf{559}     & \textbf{1100}   & \textbf{544} \\
		\hline
		\multicolumn{1}{|c|}{\multirow{8}[2]{*}{\textbf{Python}}} & \textbf{dataframe} & 262   & 160   & 100   & 171 \\
		\cline{2-6}          & \textbf{pandas} & 1121  & 356   & 100   & 121 \\
		\cline{2-6}          & \textbf{numpy} & 1115  & 249   & 100   & 121 \\
		\cline{2-6}          & \textbf{scipy} & 273   & 60    & 100   & 114 \\
		\cline{2-6}          & \textbf{scikit-learn} & 227   & 93    & 100   & 84 \\
		\cline{2-6}          & \textbf{matplotlib} & 668   & 239   & 100   & 89 \\
		\cline{2-6}          & \textbf{pyqt} & 103   & 53    & 100   & 177 \\
		\cline{2-6}          & \textbf{Total} & \textbf{3769} & \textbf{1210} & \textbf{700} & \textbf{877} \\
		\hline
	\end{tabular}%
	\label{tbl:dataset}%
\end{table}%

%% file: bugtypeClassification.tex
\begin{table}[htbp]
	\fontsize{7.5}{7.5}\selectfont
  \centering
\setlength{\belowcaptionskip}{.01cm}
  \caption{Classification of Bugs}
    \begin{tabular}{|p{0.2em}|p{7.25em}|p{21.11em}|}
   \hline
    \rowcolor[rgb]{ .718,  .718,  .718} \multicolumn{2}{|l|}{\textbf{Bug Types}} & \textbf{Description} \\
    \hline
    \multicolumn{2}{|p{8.11em}|}{\textbf{ Interface, Integration, and System (IIS) Bugs}} & These bugs occur due to the external and internal interface, operating systems (OS), and resource management problems, different versions of libraries, bugs in the libraries themselves, and lack of inter-library compatibility.  \\
    \hline
   \multicolumn{2}{|p{8.11em}|}{\textbf{Coding Bug (CB)}} & Occurs due to the syntax error.  \\
    \hline
    \multicolumn{2}{|p{8.11em}|}{\textbf{Data Bug (DB)}} & Happens due to the wrong format of the input data at early stages of data science pipelines. These bugs may lead to crash or incorrect functionality. \\
    \hline
 \multirow{5}[35]{*}{\rotatebox[origin=c]{90}{\textbf{Structural Bug (SB)}}}
      
     & \textbf{Control Flow Bugs (CFB)} & Occurs by mistakes in the control flow of the data science program such as incorrect loops, incorrect if-else, the wrong indentation in Python, and unreachable code. Common effects of this bug are crash, incorrect functionality, or data corruption.  \\
\cline{2-3}          & \textbf{Data Flow Bugs (DFB)} & This is similar to Data bugs; however, it appears at the later stages of data science pipeline. Shape mismatch and type mismatch in the data wrangling may cause these bugs. \\
\cline{2-3}          & 
\textbf{Initialization Bug (IB)} & 
Happens when a developer doesn't initialize variables or functions' parameters prior to using them and may result in a crash. \\
\cline{2-3}          & \textbf{Logic Bug (LB)} & Occurs due to the confusion with statistical background, lack of domain knowledge, or an issue in the algorithm or programming. The effects are crash, Hang, or Incorrect functionality that may require  changing few lines of code or using another library. \\
\cline{2-3}          & 
\textbf{Processing Bug (PB)} & 
This happens when one API or function is misused and easier to fix as it does not require many changes. \\
    \hline
    \end{tabular}%
  \label{tab:bugtypes}%
\end{table}%

%% file: rootcauseClassification.tex
\begin{table}[htbp]
	\fontsize{7.5}{7.5}\selectfont
	\centering
	\setlength{\belowcaptionskip}{.01cm}
	\caption{Classification of Root Causes of Bugs}
    \begin{tabular}{|p{0.3em}|p{7.63em}|p{20.11em}|} 
    	\hline
    	\rowcolor[rgb]{ .851,  .851,  .851} \multicolumn{2}{|l|}{\textbf{Root Causes}} & \cellcolor[rgb]{ .718,  .718,  .718}\textbf{Description} \\
    	\hline
    	{\multirow{4}[20]{*} {\rotatebox[origin=c]{90}{\textbf{Packge/Library}}}} & {\textbf{Absence of inter API compatibiltiy (AAC)}} & Inconsistency of the combination of two different kinds of libraries leads to a crash in data analytics programs.  \\
    	\cline{2-3}          & {\textbf{Package/Library Misselection/Conflicts (PLMC)}} & Wrong choices of packages or libraries, conflicts between packages, and missing package installation or updating \\
    	\cline{2-3}          & {\textbf{API Change (APIC)}} & New packages may cause previously working code to crash. There are bugs in the libraries that developers are not aware of them. \\
    	\cline{2-3}          & {\textbf{Wrong Documentation (WD)}} & Wrong information in the specification and documentation of APIs or packages may confuse developers and results in bugs. \\
    	\hline
    	\cline{2-3}    {\multirow{4}[25]{*}{\rotatebox[origin=c]{90}{\textbf{API}}}} & 
    	{\textbf{Shape Mismatch (SM)}} & This may happen in the following steps: apply transformation on data frames, or modify, arrange, summarize the rows and columns. \\
    	\cline{2-3}          & {\textbf{Type Mismatch (TM)}} & Both R and Python don't have static type checking, and the absence of this feature might result in crash bugs or incorrect functionality. \\
    	\cline{2-3}          & {\textbf{Incorrect Data Analysis Parameters (IDAP)}} & Wrong parameters are the major cause of crash and incorrect functionality in data analytics programs. Developers make wrong choices in the parameters at different data analysis steps. \\
    	\cline{2-3}          & {\textbf{Misuse Required Parameter (MRP)}} & An API misuse by providing the wrong number of required parameters. Unlike IDAP, the parameter is required in this category. \\
    	\hline
    	\multicolumn{2}{|p{7.63em}|}
    	{\textbf{Confusion with Data Analysis (CDA)}} & Confusion with different steps of data analysis that results in incorrect functionality or crash. To fix these bugs, usually developers need to change many lines of codes. \\
    	\hline
    	\multicolumn{2}{|p{7.63em}|}
    	{\textbf{Data Analysis Inefficiency (DAI)}} & Data analysis pipelines might not be efficient and cause bad performance. This requires an efficient implementation of the program. DAI causes logic bugs (LB) because of changes in the library, packages during data analytics steps. \\
    	\hline
    	\multicolumn{2}{|p{7.63em}|}
    	{\textbf{Others}} & These are coding or syntax error along with operating system-related bugs, etc. \\
    	\hline
    \end{tabular}%
	\label{tab:rootcauseClassificationTable}%
\end{table}%

%% file: effectsClassification.tex
\begin{table}[htbp]
	\fontsize{7.5}{7.5}\selectfont
  \centering
  \caption{Classification of Effects of Bugs}
    \begin{tabular}{|p{6.335em}|p{23.335em}|}
    \hline
    \rowcolor[rgb]{ .851,  .851,  .851} 
    \textbf{Effect} & \cellcolor[rgb]{ .718,  .718,  .718}\textbf{Description} \\
    \hline
    \textbf{Incorrect Functionality (IF)} & This effect happens when a program provides unexpected results, e.g., wrong chart, without any compilation error or warning. \\
    \hline
    \textbf{Crash} & 
    The most frequent effect in data analytics programs is crash. Programs provide errors and stop working. \\
    \hline
    \textbf{Bad Performance (BP)} & Even though there might not be any wrong usages of libraries by data scientists, there might be a bad or poor performance in data analytics programs. \\
    \hline
    \textbf{Data Corruption (DC)} & This is due to unexpected outputs or data corruption. Root causes are Confusion with Data Analysis (CDA) or Incorrect Data Analysis Parameter (IDAP). \\
    \hline
    \textbf{Hang} & 
    Some data analytics programs run for a long time without any output that lead developers to terminate them. \\
    \hline
    \textbf{Memory Out of Bound} & This happens due to the huge input file, inefficient implementation of data manipulation, etc.  \\
    \hline
    \textbf{Unknown (UN)} & 
    Symptoms could not be obtained on some \gh fix commits without documentation, title, or issue link. \\
    \hline
    \end{tabular}%
  \label{tab:effectClassification}%
\end{table}%

%% file: rq1.tex
\section{Understanding Bug Type, Root Cause, Effect and their relationship}
\label{sec:rq1to3}
\subsection{RQ1: Classes of Frequent Bugs}
\label{sec:rq1}
In this section, we address \textbf{RQ1}, describing the most frequent bugs of data analytics programs in R and Python. 
For each bug type, first, we present the \sof dataset findings and 
then compare these observations with the \gh dataset.
\figref{fig:typesRPython} shows the distribution of bug types in
 \sof and \gh datasets.
We categorize comparable R and Python packages based on similar functionality.
We used the Wilcoxon-Mann-Whitney test, designed when normality is not assessed,  to check whether the distribution of two datasets from \sof and \gh are significantly different or not.  
Our null hypothesis, $H_0$, is the distribution of the two datasets are the same, 
meaning they follow the same patterns. 
We failed to reject the null hypothesis for both languages in the \sof and \gh dataset. All the follow-up analysis in the results section were conducted on the \sof dataset due to the high-quality posts of the bug and fix. We have not included the \gh dataset, as it did not contain an explanation in the \gh commits or bug issues on the respective repositories.

\input{bugFigRPython.tex}

\subsubsection{Logic Bugs (LB)}
\label{subsec:LB}

LB bugs appear in the {\bf 46.87}\% and {\bf 51.32}\% of \sof dataset of data analytics programs written in R and Python, respectively. Similarly, structural logic bugs are the most frequent bugs in the \gh dataset for {\bf 41}\% and {\bf 39}\% of R and Python programs, respectively. 
These kinds of bugs are usually due to a lack of knowledge or experience with data analysis. 
We found that logic bugs for visualization and mathematical functions, and ML applications follow the same patterns. 
Similarly, for data manipulation and data acquisition categories, 
LB is also the most common bug in the \gh dataset.
We have conducted a follow-up analysis on the sample of 100 LB and found that about 80\% of 
these bugs in both languages occurred due to confusion of implementing tasks. For example, data cleaning such as drop or input missing values, data transformation such as variable or column transformation, and reshaping the data, for instance, sorting, grouping by, aggregation, join, and mutation. 



\subsubsection{Data Bugs (DB)}
\finding{Python exhibits significantly larger number of Data bugs compared to R.}

Python shows 12\% Data Bugs while R exhibits only 1.4\%.
From our observation, handling wrong data format issues needs more effort in Python. Developers tend to ask more questions about the early stages of data analytics programs. R provides in-built features for data acquisition, while Python developers use \pandas and \numpy with several different optional arguments to load data. 
\input{DBPyRExample.tex}

For example, the above code snippet from open Git issue ~\cite{gitissueCSVnrows} in \textit{pandas-dev} reveals how the python-parser provides unexpected output in case of reading a headerless CSV file with \textit{nrows=1} or \textit{nrows=2}. 
When a developer uses \textit{read\_csv()} with a specific number of \textit{nrows}, the \textit{DataFrame} is returned correctly with \textit{nrows} long, but the file pointer behaves incorrectly that prevents the file reading from the desired point. 
However, this kind of issue~\cite{gitissueCSVr} with built-in \textit{read.csv()} function does not occur in R. 
Such kind of difficulty for reading header-less CSV files is already addressed in R~\cite{dbissue1, dbissue2}, which leads to the following implication:

\textbf{Implication:} \textit{We believe that the researchers could be informed by the API design of R's built-in functionalities for data acquisition to enrich API design in Python packages.}


\subsubsection{Data Flow Bugs (DFB)}
\finding{Data Flow Bugs appear more in R programs than Python for data manipulation tasks.}

DFB bugs appear in the {\bf 16}\% and {\bf 10}\% of data analytics programs written in R and Python respectively. 
Since R uses more pipes in data wrangling tasks, the intermediate results are not stored explicitly. 
For example, while using \textit{dplyr} package of R for grouped operations that result in length $\ne$ 1 or length(group) throws an error as mentioned in the \sof post~\cite{so21737815}.

\input{examples/DFBPyR}

Therefore, R developers have less control over data types and shapes, while the explicit representation of intermediate results in Python leads to less DFB in Python compared to R. 
We conducted a deeper analysis on the Data Flow Bugs dataset and found out that
R developers have used about 42\% pipes, while Python developers only used 5\% pipes or chain operators.
For example, a developer explicitly defined \texttt{items\_counts} variable before calling a \texttt{max()} function in the next line~\cite{so15138973}. 
We further analyzed and found that Shape Mismatch (SM) causes {\bf7.16}\%  and {\bf 4.96}\% of total bugs in R and Python data analytics programs.
This is due to the less control and observation of implicit results in
the data science pipeline in R. 
We identified several Git issues that request improvement in R packages for enhancing debugging facilities regarding data flow issues such as~\cite{dfbissue1, dfbissue2, dfbissue3}.

\textbf{Implication:} 
\textit{This seems to indicate that R developers can benefit from debugging facilities to help examine values at intermediate points in data analytics programs, e.g., variable watch.}

\subsubsection{ Interface, Integration, and System (IIS) Bugs}

\finding{On average there are more IIS bugs in R than in Python.}
IIS bugs appear on average {\bf 16.46}\% and {\bf 14.96}\%  in R and Python programs respectively. Even though Python has more packages, the dependencies between R packages are higher~\cite{decan2016topology} which results in a similar number of IIS bugs. Other sources of IIS bugs are more general among R and Python developers, such as changing operating systems or changing underlying graphics library \textit{ggplot2}~\cite{so31717850}.
\input{examples/IISpy}
\input{examples/IISr}
As the error message implies, there is a dependency to the `Rcpp' package. Wang~\etal~\cite{wang2018dependency} built an automatic detection tool to address the dependency conflict for Java projects. A similar dependency analysis tool would help practitioners alleviate these issues.
Additionally, we have identified several Git issues~\cite{plmc1, plmc2, plmc3, plmc4, plmc5} that request the solutions for dependency issues in R packages. However, dependency management tools like Packrat~\cite{ushey2015packrat} have been deployed for some R packages to help with dependency issues.
\textbf{Implication:}
\textit{These findings along with prior work on inferring environment dependencies in Python packages~\cite{horton2019dockerizeme} highlight the need for similar research for R.}

\subsubsection{Initialization Bugs (IB)}

\finding{Developers face more IB bugs in R than Python for programs written for similar tasks.}
Initialization Bugs appear on average {\bf 10.91}\% and {\bf 3.72}\%  in R and Python programs respectively. 
Due to the functional paradigm of R in which outputs are provided implicitly as an input of the next function, any simple mistake in the initialization of a function causes the next steps of data manipulation or visualization step to fail.  
 \input{examples/IBpy}
\textbf{Implication:} 
\textit{Researchers could utilize our benchmark to design and develop an automated data initialization tool for initializing variables with their proper types and values. For example, when a developer does not explicitly initialize a variable, the tool would recommend its appropriate type and value. This tool would be useful when the program executes a statement that started in the middle of a code block, but the local variables were not initialized within that block.}



%% file: bugFigRPython.tex
\begin{figure}[h!t]
		\centering
	\subfloat[\textbf{Stack Overflow}]{{\includegraphics[scale= .46, trim={0cm 1cm 0cm 0cm},clip]{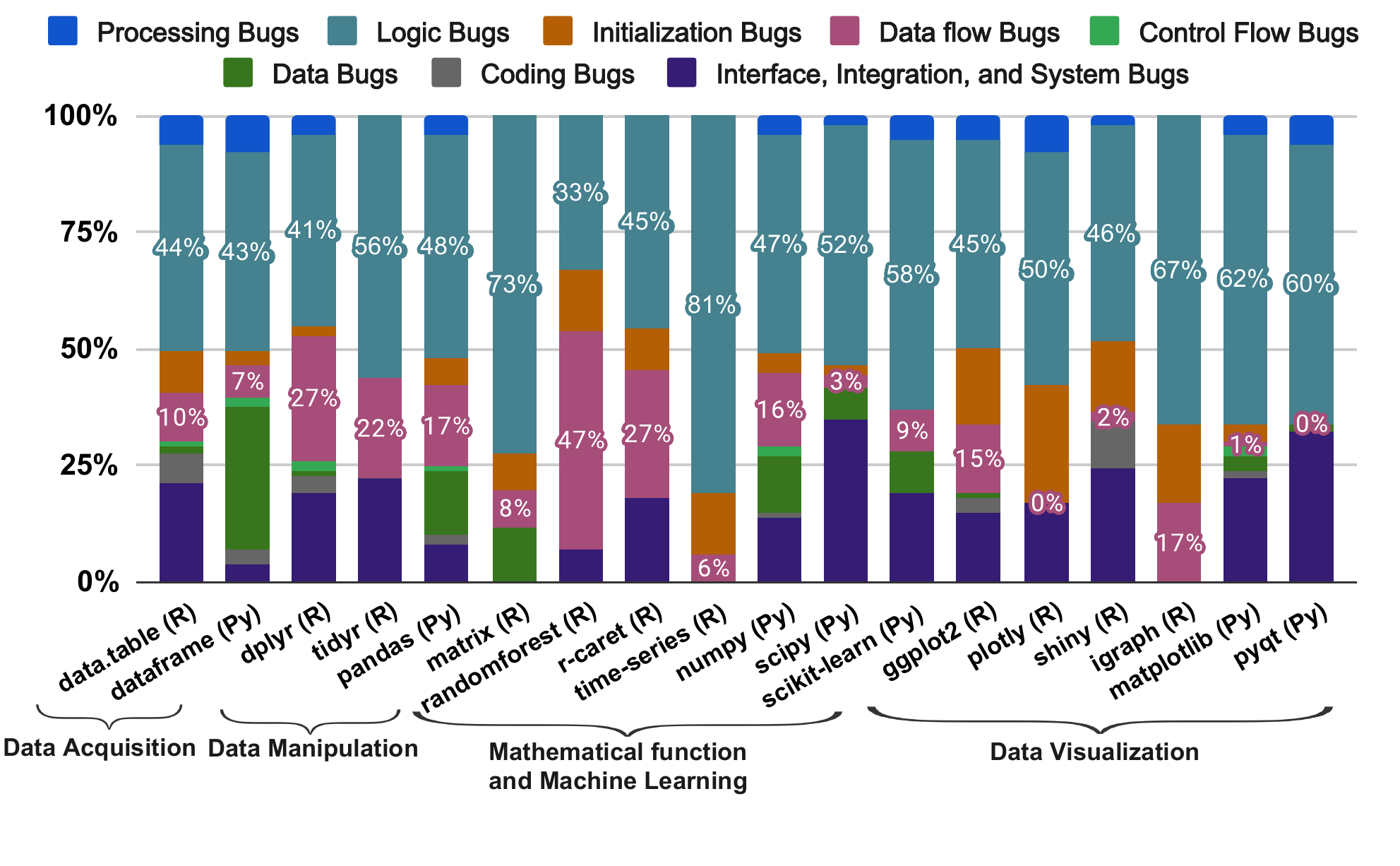}}}%
	\qquad
	\subfloat[\textbf{GitHub}]{{\includegraphics[scale= .46,trim={0cm 2.2cm 0cm 0cm},clip]{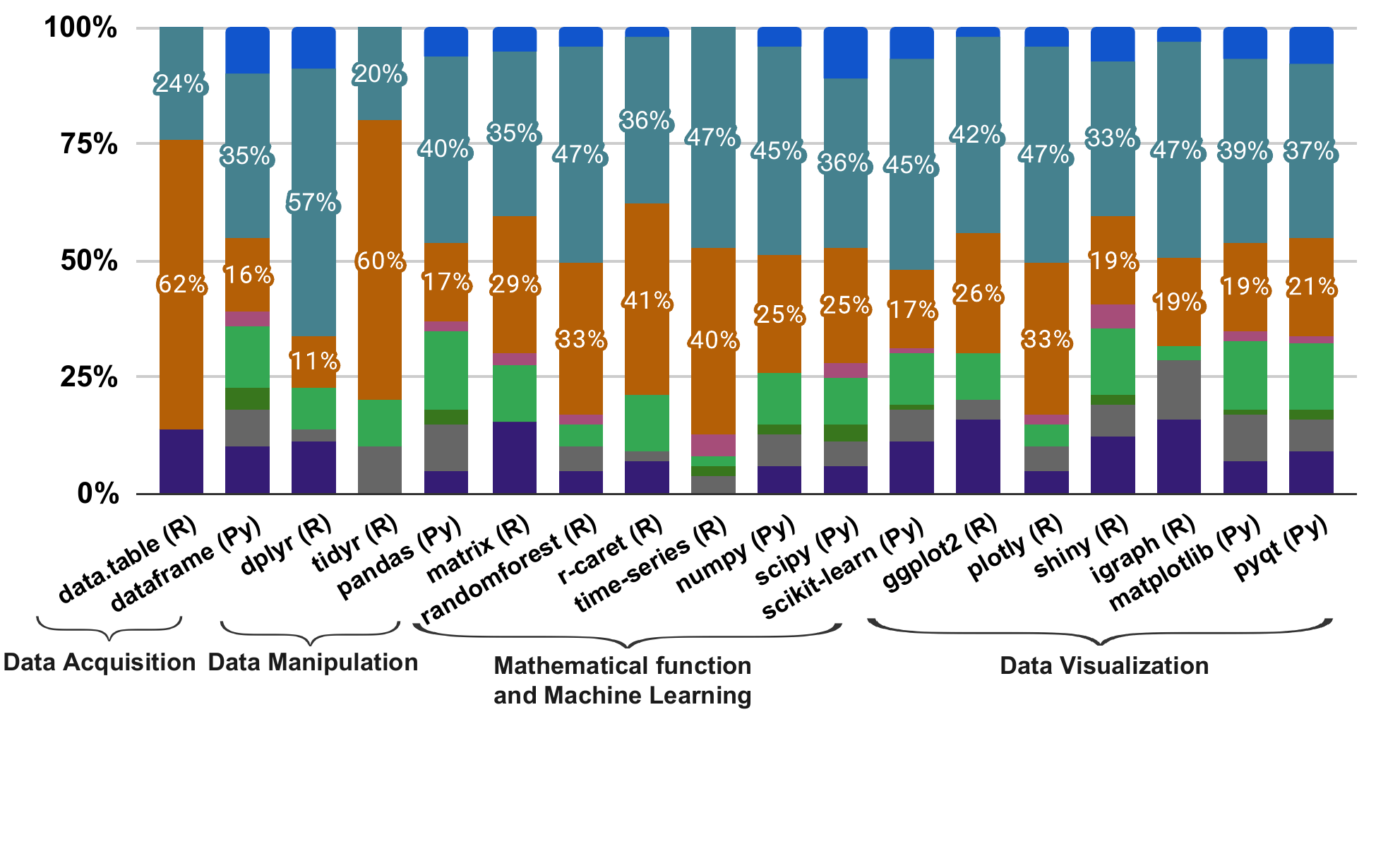}}}%
	\caption{Distribution of bug types across Python and R packages (\textit{2 most frequent bug types are highlighted (\%)}}   
	\label{fig:typesRPython}
\end{figure}

%% file: DBPyRExample.tex
\begin{figure}[h]
	\includegraphics[width=\linewidth,trim={0cm 11.1cm 0cm 0.8cm},clip]{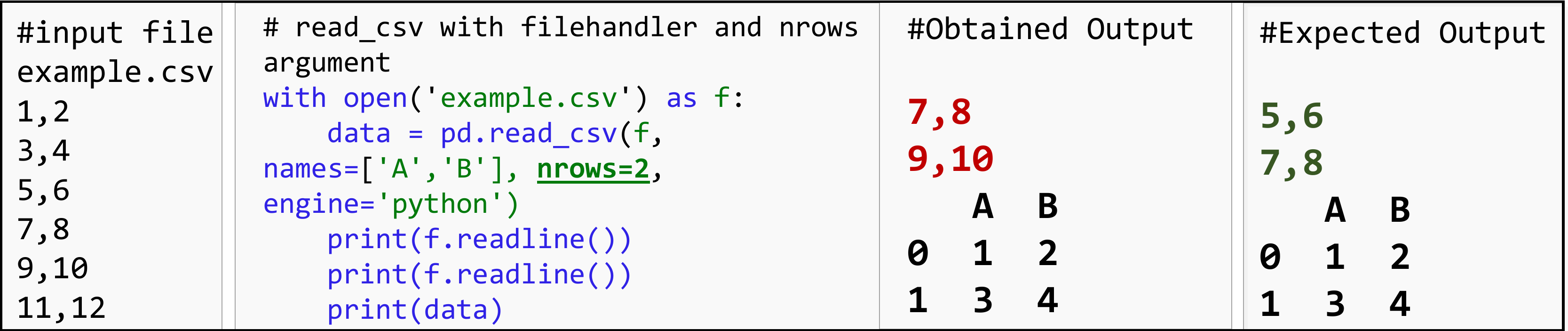}
	\label{fig:evolution}
	\vspace{-0.8cm}
\end{figure}

%% file: examples/DFBPyR.tex
\begin{lstlisting}[basicstyle=\fontsize{6}{6}\ttfamily,language = R, numbers=left]
df = data.frame(a = 1:4, b = 1:2)
df %>% group_by(b) %>% summarise(rep(a[1], 3))
\end{lstlisting}

A \textit{do} operator can resolve the issue.
\begin{lstlisting}[basicstyle=\fontsize{6}{6}\ttfamily,language = R, numbers=left]
df %>% group_by(b) %>% do(data.frame(a = rep(.$a[1], 3)))
\end{lstlisting}

%% file: examples/IISpy.tex
For instance, a developer wants to read a .xlsx with \pandas, but gets the following error~\cite{so48066517}:
\begin{lstlisting}[basicstyle=\fontsize{6}{6}\ttfamily,language = python, numbers=left]
data = pd.read_excel(low_memory=False, io="DataAnalysis1/temp1.xlsx").fillna(value=0) 
ImportError: Install xlrd >= 0.9.0 for Excel support
\end{lstlisting}
The solution would be to install `xlrd' as mentioned by an expert with high reputation on \sof. 


%% file: examples/IISr.tex
\begin{lstlisting}[basicstyle=\fontsize{6}{6}\ttfamily,language = R, numbers=left]
> library(ggplot2)
Error in loadNamespace(j <- i[[1L]], c(lib.loc, .libPaths()), versionCheck = vI[[j]]) :
there is no package called 'Rcpp'
Error: package or namespace load failed for 'ggplot2'
\end{lstlisting}


%% file: examples/IBpy.tex
For example, in a the data cleaning process, a developer may get initialization bug while querying for $NaN$ and other names in \pandas using following code~\cite{so26535563}. According to an expert developer mentioned in that \sof post, using $@nan$ instead of $NaN$ can solve the issue.
\begin{lstlisting}[basicstyle=\fontsize{6}{6}\ttfamily,language = python, numbers=left]
df.query( '(value < 10) or (value == NaN)' )
\end{lstlisting} 

%% file: rq2.tex

\subsection{RQ2: Frequent Root Causes}
\label{sec:rq2}
In this section, we answer \textbf{RQ2} and present our findings. For each root cause, first, we present the \sof dataset finding and then compare with the \gh dataset.
\figref{fig:rootcausesPythonR} shows the distribution of root causes in \sof and \gh datasets for each package in R and Python.
\input{causeFigRvPython}

\subsubsection{Confusion with Data Analysis (CDA)}

CDA is the root cause of  {\bf 44.19}\% and {\bf 49.71}\% of bugs in the R and Python programs, respectively in \sof dataset. For example, an R developer wants to run a statistical t-test over multiple variables against the same categorical variable by the following code but got an error~\cite{so26244321}. This problem requires knowledge of statistical testing. Here, $vs$ is either 0 or 1 (group), and it needs a subset of data. The following patch contains the bug (line 2) and fix (line 3) suggested by an expert developer.

\input{examples/CDAr}

In another \sof post, a Python developer had a problem getting the field ~\cite{19202093}.
The user grouped the dataframe into two columns and asked how to access the `name' field of the resulting median. The user was confused about which parameter were needed to pass. There are many ways to solve this problem, but the best solution is to pass the index ‘name’ using get\_level\_values() API.
\begin{lstlisting}[basicstyle=\fontsize{6}{6}\ttfamily,language = diff, numbers=left]
df = pd.DataFrame({'a': [1, 1, 3],
                   'b': [4.0, 5.5, 6.0],
                   'c': [7L, 8L, 9L],
                   'name': ['hello', 'hello', 'foo']})
- df.groupby(['a', 'name']).median()
+ df.groupby(["a", "name"]).median().index.get_level_values('name')

\end{lstlisting}


These kinds of bugs are caused due to the confusion with data analytics steps. CDA is the root cause of 77\% of logic bugs (LB) in Python and 85\% of LB in R programs. For example, the logic bug in Section ~\ref{sec:rq1}, happened due to the confusion with the data manipulation task that changed a few steps of the pipeline. Similarly, in \gh dataset, CDA is the common root cause of 
{\bf 50.18}\% and {\bf 63.81}\% of total bugs in R and Python respectively. CDA is the main root cause of DB, which contributes 42\%. We conducted a deeper analysis of 100 samples of bugs in this category and found that 26\% and 40\% of posts in R and Python had bad code smell~\cite{fowler2018refactoring}, i.e., an indication of a deeper problem in the code, including long parameter list, too many returns, long line, especially for the statistical/mathematical expressions.

\textbf{Implication:} 
\textit{An empirical study on design flaws such as bad code smell would help the SE community to better understand prevalent design flaws in data analytics applications and develop code smell detection tools such as tsDetect~\cite{tsDetect}. This also suggests that educators should incorporate more SE concepts into their curriculum. Similar to the recurring bugs problem in SE~\cite{gao2015fixing}, we need an empirical analysis to detect such patterns of confusion in data analytics programs.}


 
\subsubsection{Incorrect Data Analysis Parameters (IDAP)}

\finding{ IDAP on average causes a similar number of bugs in R and Python.}

On average IDAP is the root cause of  {\bf 9.66}\% and {\bf 10.26}\%  of bugs in R and Python programs respectively.
\input{examples/IDAPpy} 
Similarly, IDAP in \gh dataset is the second common root cause of bugs, and it causes {\bf33}\% and {\bf 20}\%  bugs in R and Python, respectively. Further investigation has revealed that 15\% and 25\% of the data bugs, respectively, in Python and R programs are caused by IDAP. IDAP is also the main root cause for IB in both R and Python programs, contributing to approximately 82\% of bugs. Incorrect Data Analysis Parameters (IDAP) is easier in GH to detect as the fix contains only one argument change. If corresponding libraries enhanced the APIs with default parameter initialization, these bugs would not occur.

\textbf{Implication:} 
\textit{Data bugs and data flow bugs caused by IDAP could be addressed by an automatic tool that utilizes the specification of each library. Similar automatic parameter recommendation tools are needed, e.g., parameter usage for Java developers~\etal~\cite{zhang2012automatic} and mining API function calls and usage patterns~\cite{focus}.}

%


\subsubsection{API Changes (APIC)}

\finding{API and Package changes lead to significantly more bugs in R compare to Python. }

APIC causes {\bf9.48}\%  of bugs in R while it causes  {\bf 3.39}\% of Python bugs. This is surprising due to the huge number of Python packages. According to the R documentation~\cite{rDocumentation} website at the time of writing this paper, there are 17,984 packages on \CRAN, 
\BioConductor and \gh, while Python's \textit{PyPI} has 192,858 packages. Libraries in Python and  R tend to update frequently and lead to a crash. Library developers tend to reuse already available packages, and these changes propagate to dependent packages. These dependencies between packages are more severe in R programs~\cite{decan2016topology}, and the number of isolated Python libraries that are not dependent on others is much higher. This suggests that there is a need for a better dependency analysis tool for R programs.

We further analyzed eight years of closed issues on the selected R and Python packages. We found that Python API designers utilized significantly more API deprecation strategy than R API designers.  \figref{fig:deprecation} shows that the numbers of closed issues related to deprecation in Python packages (\pandas, \numpy, \matplotlib) are higher than in R packages (\ggplot, \dplyr, \tidyr). Upon further investigation of those closed issues, we found that the difference in the number of deprecation-related issues between Python and R packages could be due to various factors. Those factors are library popularity, development rate, policies for deprecating APIs, number of users and contributors, etc.

\textbf{Implication:}
\textit{This suggests that library developers of R packages should focus on improving documentation and deprecation strategies.}

\input{deprecateevolution.tex}

\subsubsection{Package library Mis-selection/Package Conflicts (PLMC)}
\finding{PLMC causes more bugs in Python than R programs.}
PLMC causes {\bf5.01}\%  of total bugs in R while it causes  {\bf 7.69}\% of Python bugs.
As an example, consider the listing below where an R developer is facing bugs in a
\sof post~\cite{so30562819}:
\input{examples/PLMCr}
Considering the sheer number of packages that are available in these languages,
we have found the percentage of bugs attributed to PLMC smaller than expected. 

\textbf{Implication:}
\textit{
There is a need for dependency-checking tools to monitor dependency conflicts for the R similar to Watchman~\cite{wang2020watchman}, which targets the Python library ecosystem.
}


%% file: causeFigRvPython.tex
\begin{figure}[h!t]
		\centering
	\subfloat[\textbf{Stack Overflow}]{{\includegraphics[scale= .46, trim={0cm 1cm 0cm 0cm},clip]{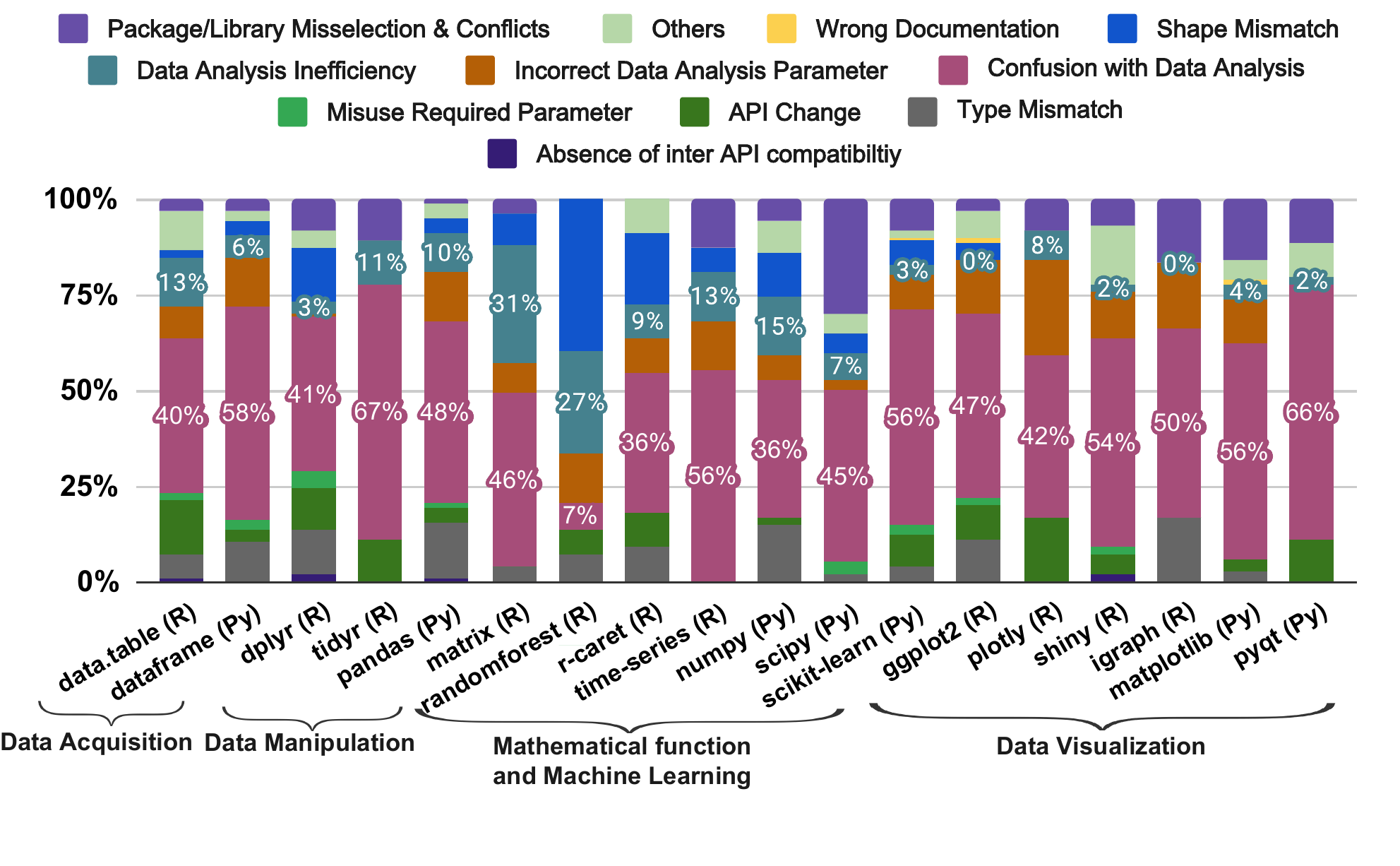}}}%
	\qquad
	\subfloat[\textbf{GitHub}]{{\includegraphics[scale= .46,trim={0cm 2.5cm 0cm 0cm},clip]{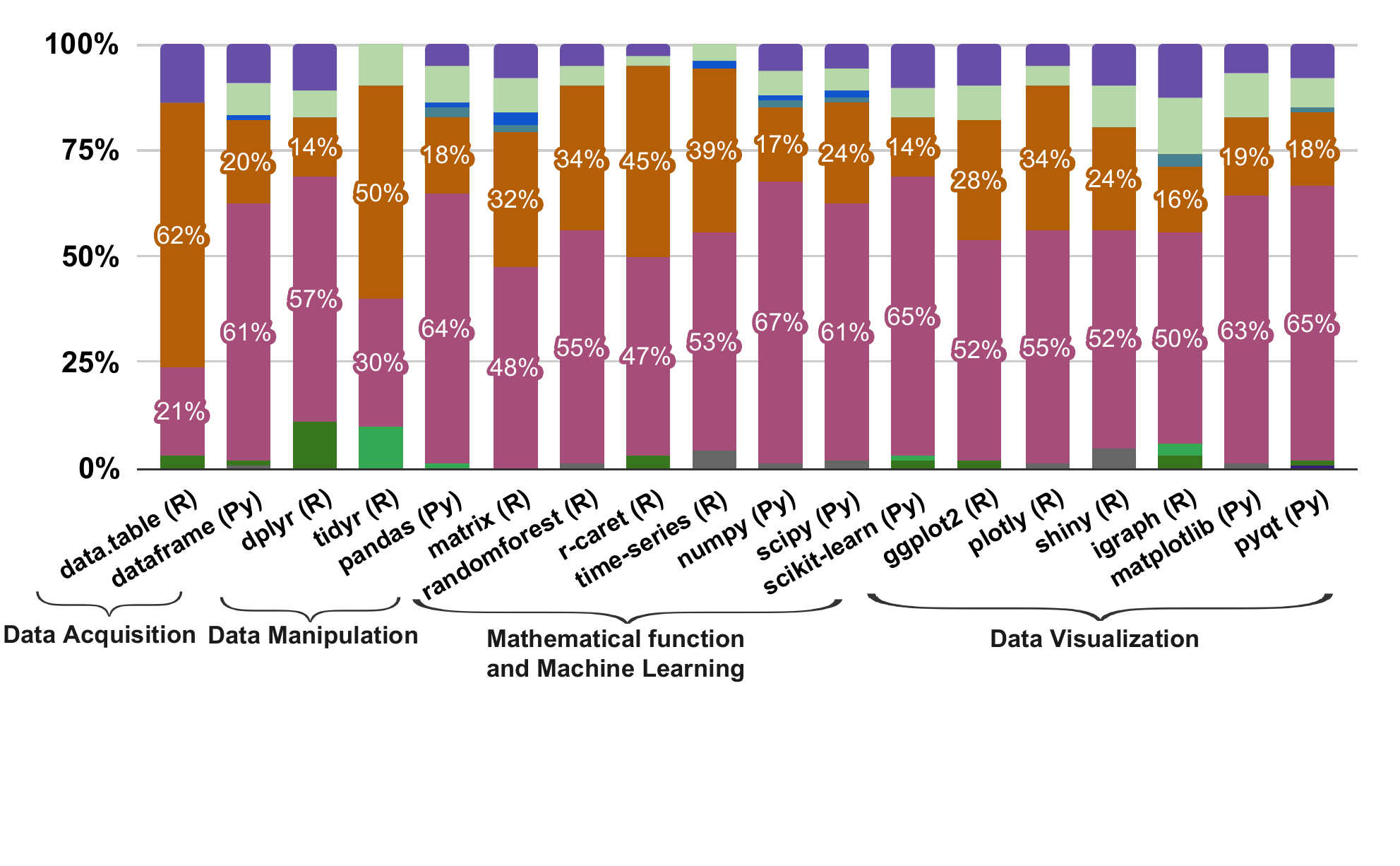}}}%
	\caption{Distribution of root causes of bugs across Python and R packages (\textit{2 most frequent root causes are highlighted})}   
	\label{fig:rootcausesPythonR}
\end{figure}

%% file: examples/CDAr.tex



\begin{lstlisting}[basicstyle=\fontsize{6}{6}\ttfamily,language = diff, numbers=left]
library(dplyr)    
- mtcars %>% summarise_each_(funs = (t.test(. ~ vs))$p.value, vars = disp:qsec)
+ mtcars %>% summarise_each(funs(t.test(.[vs == 0], .[vs == 1])$p.value), vars = disp:qsec)
\end{lstlisting}

%% file: examples/IDAPpy.tex
As an example of IDAP bug consider the following code where 
a Python developer wants to load a csv file~\cite{so9652832}:
\begin{lstlisting}[basicstyle=\fontsize{6}{6}\ttfamily,language = python, numbers=left]
DataFrame.from_csv('c:/~/trainSetRel3.txt', sep='\t')
\end{lstlisting}

Developer encountered a \textit{PandasError}. To fix this bug, developer needs to call the API with to add \textit{header=0} parameter. 

%% file: deprecateevolution.tex
\begin{figure}[h]
\centering
	\includegraphics[width=3.2in,trim={0cm 0cm 0cm 1.5cm},clip]{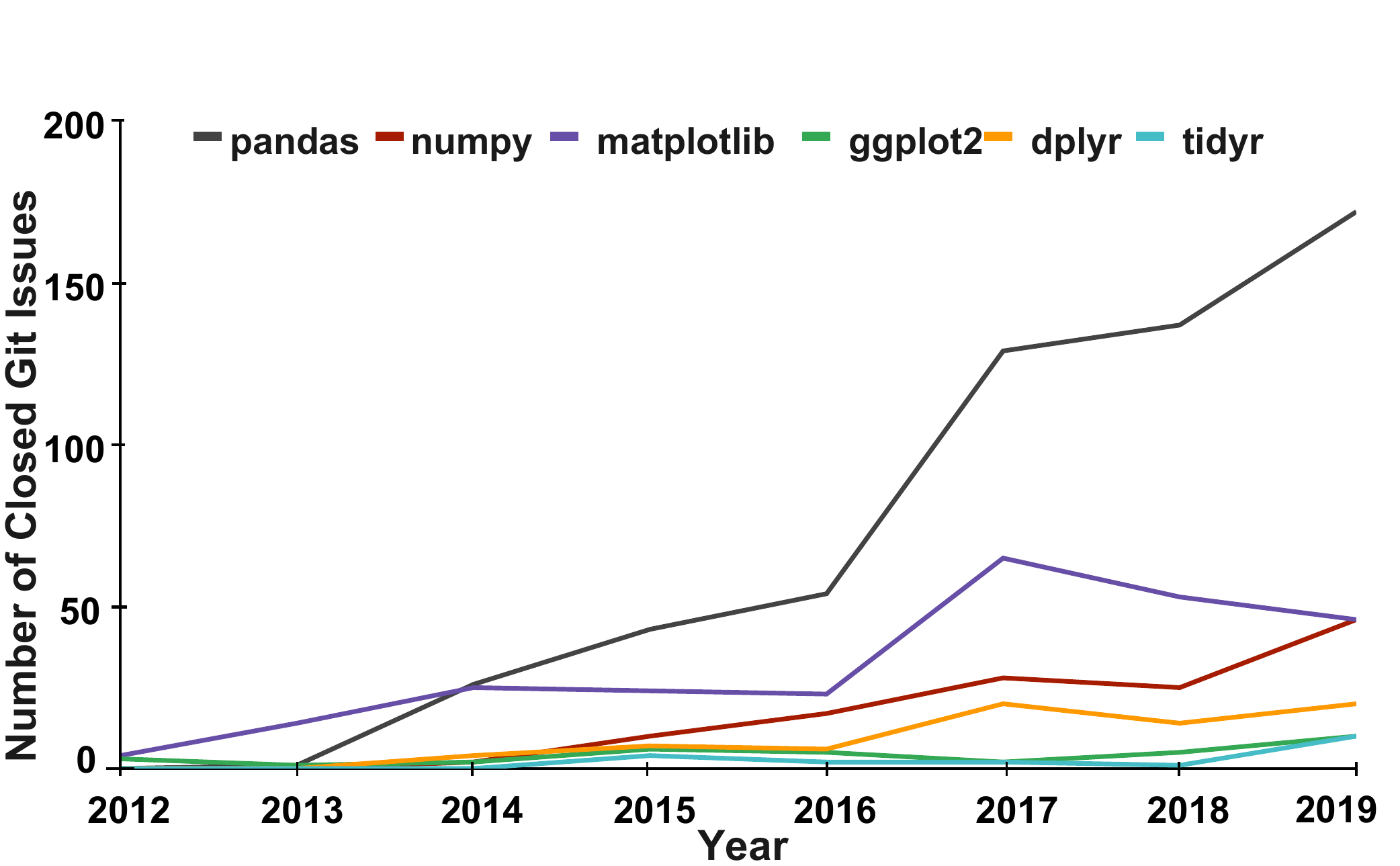}
	\caption{Deprecation history of Python packages (pandas, numpy, matplotlib), and R packages (ggplot2, dplyr, tidyr)} 
	\label{fig:deprecation}
\end{figure}

%% file: examples/PLMCr.tex
\begin{lstlisting}[basicstyle=\fontsize{6}{6}\ttfamily,language = R, numbers=left]
library("dplyr")
d <- data.frame(alpha=1:3, beta=4:6, gamma=7:9)
rename(d, c("beta"="two", "gamma"="three"))
\end{lstlisting}



The developer faced the error message due to loading both the packages \texttt{"dplyr"} and \texttt{"plyr"} in same session that causes this bug. Similar problem could be found in another \sof post~\cite{26106146}.



%% file: rq3.tex
\subsection{RQ3: Bug Types, Root Causes and Effect Relationship}
\label{sec:rq3}
In this section, we discuss the relationship among bug types, root causes, and effects afflicting different tasks in data analytics Programs. To answer RQ3, we categorized packages into four tasks (i.e., Data Acquisition, Data Manipulation, Data Visualization, and Mathematical Function and Machine Learning) in R and Python according to similar functionality R~\cite{R} and Python~\cite{sanner1999python}.

\subsubsection{Bug Type and Root Cause relationship.}
\finding{Type Mismatch root cause occurs mostly in Data Flow bugs and the root cause Incorrect Data Analysis Parameter (IDAP) is the most frequent in Initialization bugs (IB) in both R and Python packages.}
After studying Python buggy programs, we have observed that the most common root cause for all bug types (except IB, and IIS) is Confusion with Data Analysis (CDA). For R programs, we have found that the most common root cause for all types of bugs in all data analytics tasks (except IIS) is Confusion with Data Analysis. 
To elaborate, in all tasks in R languages, we have observed that the Type Mismatch root cause occurs mainly in Data Flow bugs (87.50\%). Also, the root cause Incorrect Data Analysis Parameter (IDAP) is the most prevalent in Initialization bugs (IB) (94.44\%). In Python language, for all data analytics tasks, we have observed that the Type Mismatch is the main root cause of Data Flow bugs (54.40\%). Also, the root cause Incorrect Data Analysis Parameter (IDAP) occurs the most in Initialization bugs (IB) (82.22\%). 

\textbf{Implication:}
\textit{Based on these observations, we suggest that researchers could focus on building programming support to detect type mismatches and repair data flow bugs in R and Python. Also, developers should be aware of the initialization of parameters so that both R and Python practitioners do not get initialization bugs.}

In particular, for the data visualization task in R, the root cause Type Mismatch (TM) is the most prevalent in DB, DFB, IB, and PB bugs types. The CDA is the most common root cause in data acquisition and mathematical, machine learning tasks. MRP is the most common root cause in data manipulation tasks. 

\textbf{Implication:} \textit{Researchers should build tools to detect misuse of required parameters in data manipulation tasks for R language and utilize our dataset as a benchmark.}

In Python, CDA is the most common root cause of CFB, DB, IB, LB, and PB in data acquisition, data manipulation, mathematical function, and machine learning tasks. IDAP is the second common root cause in DB, IB, LB, and PB bugs for data visualization tasks.  

\textbf{Implication}: \textit{To help mitigate confusion in data analytics tasks, package vendors should complement the documentation and specification by following our labeled set of bugs as a reference. Furthermore, to mitigate bugs (i.e., DB, IB, LB, PB) in data visualization packages, developers should be aware of incorrect data analysis parameters pointed out by our study.}
\subsubsection{Bug types and Effect relationship.}
\finding{Crash is the most frequent bug effect in Python, caused by Coding Bug (83.33\%), Data Bug (68.46\%), Data Flow Bug (83.20\%), Interface, Integration, and System Bug (80.00\%).}
Most bug effects in R, such as Bad Performance (93.33\%), Data Corruption (74.51\%), Hang (66.67\%), and Incorrect Functionality (61.74\%) are due to Logic Bug. In Python, we found the effect Hang (80.00\%) and Memory Out of Bound (MOB) are frequent (62.50\%), but Data Corruption (38.89\%) occurred less than in R packages for Logic Bug. Also, we found the effect of Bad Performance (74.51\%) and Incorrect Functionality (70.66\%) occurs due to Logic Bug in Python packages, similar to R.
We have further investigated logic bugs and identified that they are mainly caused by confusion with data analysis steps and programming issues which might be helpful with automated code generation and completion tools. 

Therefore, code completion and generation tools for data analytics programs are needed to assist developers (e.g., OpenAI Codex~\cite{codex}, Codepilot~\cite{codepilot}). These tools can generate code snippets intended for data analytics tasks by utilizing our benchmark to address the confusion in data analytics steps and mitigate logic bugs.
On the other hand, in R, we found the Coding Bug (100.00\%), Data Bug (61.22\%), Data Flow Bug (90.91\%), Interface, Integration, and System Bug (71.43\%) results is Crash effect which we found to be similar in Python packages. Furthermore, we classified the error messages into Coding, Data, Data Flow, and IIS bugs which appeared as crash bugs which guides us to the following implications:

\textbf{Implication}: \textit{
Coding bugs in Python can be effectively detected by developing deep static/dynamic analysis tools~\cite{jupyterStatic} by using our benchmark with buggy code. Moreover, data bugs, and data flow bugs can be detected by following the design-by-contract technique~\cite{meyer1992applying}, which could be done by extending PyContracts~\cite{pycontracturl}.}
\subsubsection{Root Cause and Effect relationship.}
\finding{Most of the effects in R, such as Data Corruption (76.47\%), Incorrect Functionality (63.09\%) are due to Confusion with Data Analysis (CDA), while most of them in Python are due to Memory Out of Bound (75.00\%), and Hang (80.00\%).}

We observed that in Python packages, Data Corruption (66.67\%) and Incorrect Functionality (74.49\%) are frequent due to the root cause of CDA. For R packages, we found that the effects of Memory Out of Bound (20.00\%) and Hang (33.33\%) are less frequent than in Python packages. This suggests that Python developers suffer more from Memory Out of Bound and Hang effects in their data analytics program than in R. 
Therefore, to minimize the impact of bugs on Python and R packages, we suggest implementing a robust precondition checker tool using the required parameters to detect data corruption bugs. Such tools could be developed using our benchmark to detect data leakage in data analytics code by following prior work on static analysis for data science programs~\cite{dataleakage}. To deal with bugs resulting in Memory Out of Bound and Hang in Python programs, package developers should utilize our benchmark and resolve the memory-related and halting issues.  
\finding{Crash is the most frequent effect and is mainly caused by Misuse of Required Parameter (MRP) (92.86\%) in Python, whereas by Shape Mismatch (97.50\%) in R packages. Furthermore, most of the incorrect functionality occurs due to wrong documentation in both Python (66.67\%) and R (50.00\%) packages.}


We further analyzed and found that Misuse of Required Parameter (MRP) is due to a wrong number of parameters in the APIs of Python programs, which leads to Crash. Therefore, developers can use our benchmark to write contracts~\cite{meyer1992applying} on client code to detect the incorrect numbers of parameters and provide fix suggestions. For dealing with Shape Mismatch (SM) causing crash in R, we can replicate PyContracts~\cite{pycontracturl} for R as well to check the appropriate transformation on data frames, modification, or summarization of the rows and columns by writing a contract on respective APIs utilizing our benchmark. Additionally, for mitigating the incorrect functionality that is caused by incorrect documentation, our dataset labeled Wrong Documentation (WD) causes Incorrect Functionality (IF) could be leveraged. To mitigate bugs that cause incorrect functionality in both R and Python, our benchmark can help package vendors to improve package-wise documentation.


%% file: rq4.tex

\subsection{RQ4: Frequent Effects of Bugs}
\label{sec:rq4}

\input{effectFigRPython}


In this section, we present the answer to \textbf{RQ4} and discuss the findings for the most frequent effects of bugs in data analytics programs in R and Python. 
\figref{fig:effectPythonR} illustrates the effects of bugs using each package in R and Python in the  \sof  and \gh dataset. 
We observed that most bugs in R and Python lead to program crashes. On average {\bf 56.43}\%  and {\bf 53.72}\%  of bugs cause crashes in R and Python, respectively. Moreover, we found that R developers have encountered {\bf 5.36}\% bad performance while Python programmers have observed {\bf 8.43}\% similar effects. 
This is likely due to R's built-in statistical analysis functions, which result in fewer code lines required to perform similar tasks compared to Python. Therefore, we encouraged practitioners to implement Python versions of existing R packages with in-built statistical analysis functions to perform data analysis to mitigate crash bugs.

\textbf{Implication}: 
\textit{We encouraged researchers to design and develop bug detection and localization approaches which can leverage Crash as an effect to detect bugs in data analytics programs.}
\subsubsection{Data Corruption (DC)}

\finding{DC occurs on 9.11\% of R and on 2.98\% of Python programs.}

We have analyzed further to identify the reason why data corruption effects are more frequent in R than in Python.
We have found more bugs in R programs that are caused mainly due to the shape or type mismatch of parameters in APIs that are responsible for the bugs resulting in the data corruption effect. Another reason could be that R is used more for data manipulation, and the number of posts might be higher than Python.  

\textbf{Implication:}
\textit{Researchers can utilize our benchmark to identify which parameters in APIs are responsible for bugs resulting in Data Corruption. Also, they can design techniques to repair bugs resulting in the Data Corruption effect leveraging our dataset.}

%% file: effectFigRPython.tex
\begin{figure}[h]
		\centering
	\subfloat[\textbf{Stack Overflow}]{{\includegraphics[scale= .46, trim={0cm 0.9cm 0cm 0cm},clip]{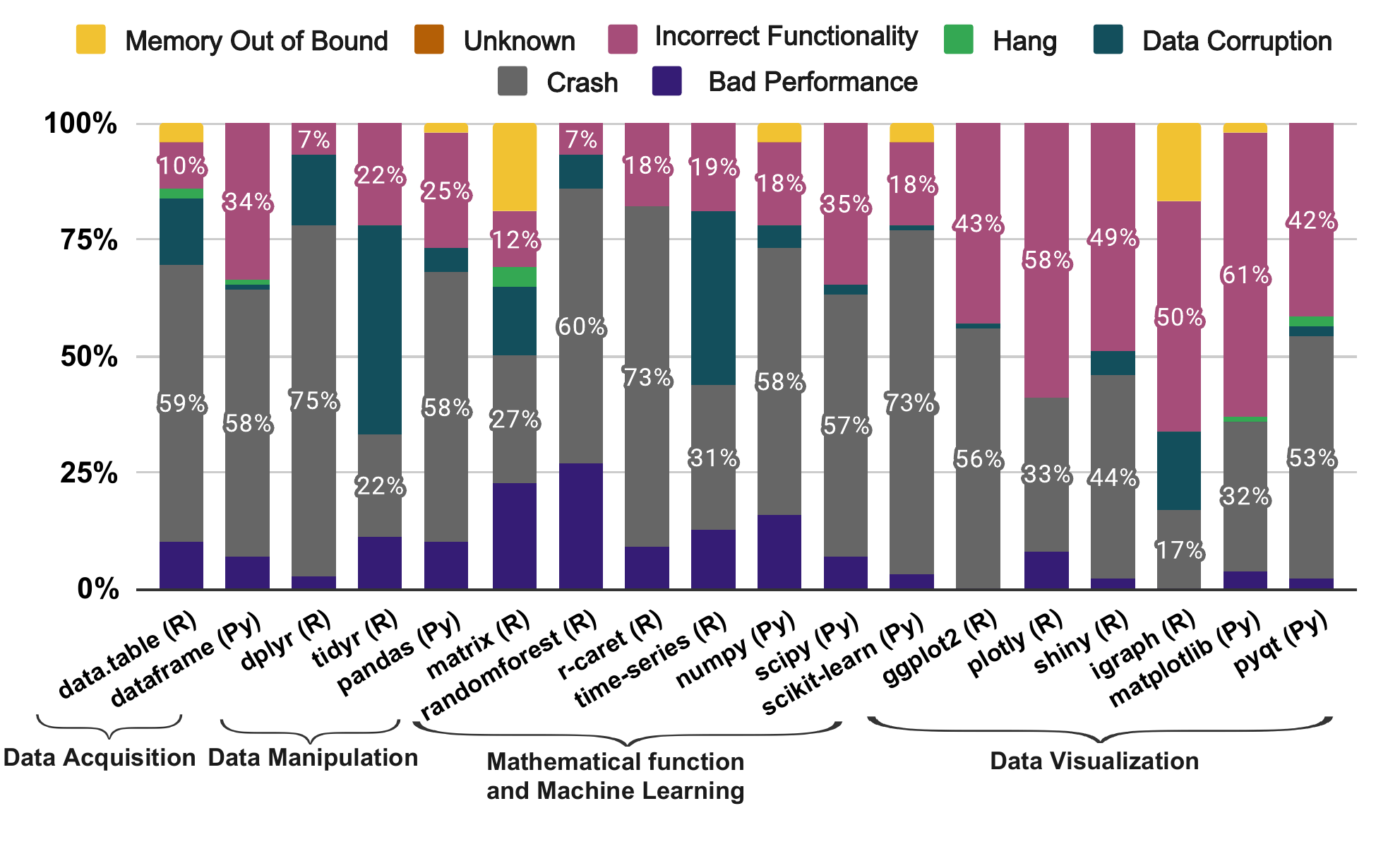}}}%
	\qquad
	\subfloat[\textbf{GitHub}]{{\includegraphics[scale= .46,trim={0cm 2.2cm 0cm 0cm},clip]{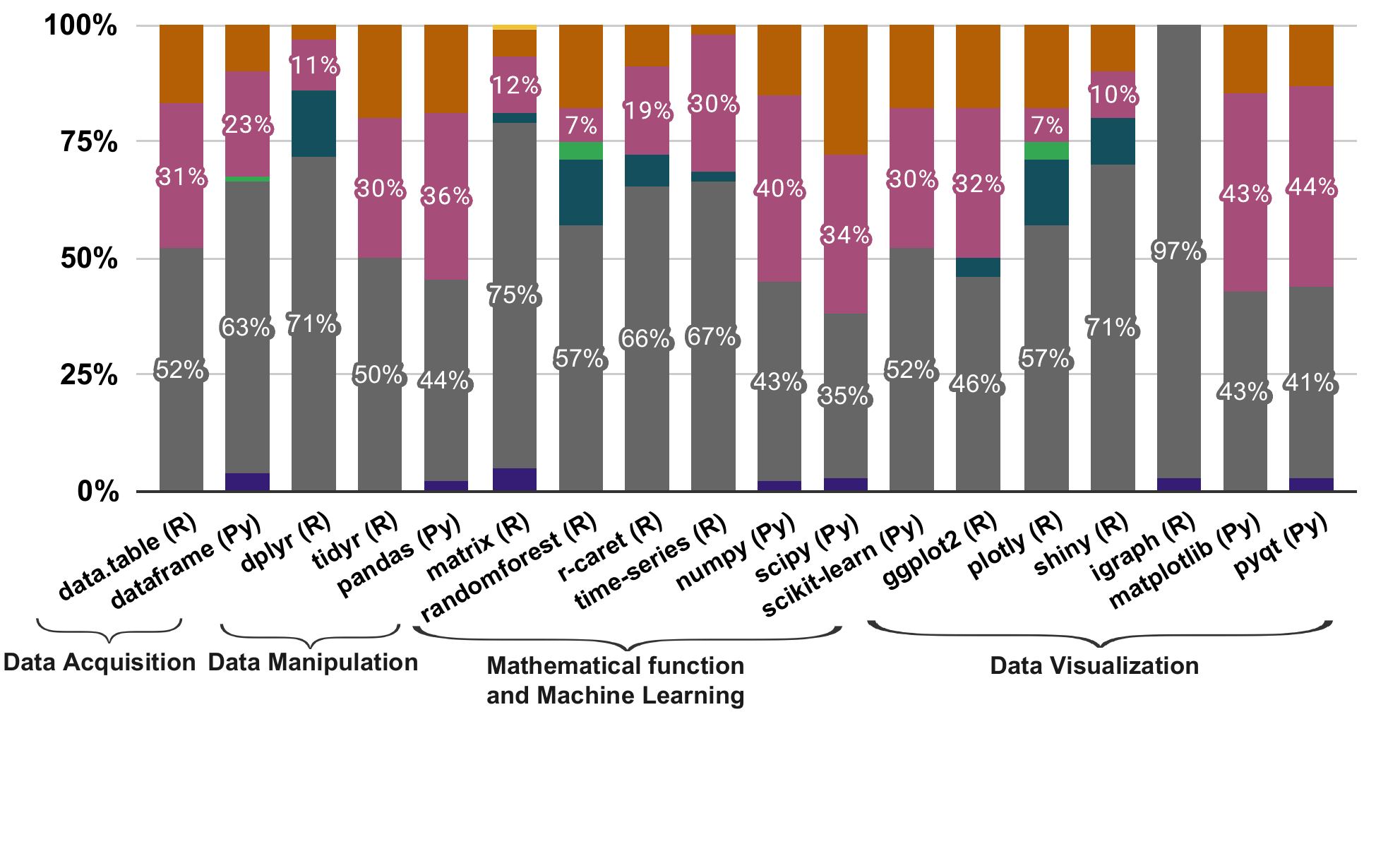}}}%
	\caption{Distribution of effects of bugs using Python and R packages (\textit{2 most frequent effects of bugs are highlighted (\%)}}   
	\label{fig:effectPythonR}
\end{figure}



%% file: rq7.tex
\section{RQ5: Commonalities of bugs types in Data Analytics Packages}
\label{sec:rq7}
\input{correlation}

In this section, we study the correlation of bug types among packages with similar functionalities in R and Python (e.g., Data Visualization, Mathematical Function, and Machine Learning). The purpose of this study is to identify whether repair and mitigation strategies in one language can be applied to other, which would be beneficial to researchers. To that end, we computed the Pearson correlation coefficients using the bugs of similar packages in R and Python. Fig.5(a) shows the \% bug types considered for R (i.e., ggplot2, plotly, shiny, igraph) and Python (i.e., matplotlib and pyqt) packages, which we use as random variables for computation. 
Our findings show that data visualization libraries are strongly correlated (\figref{fig:coefVizMath}(a)). 
Regarding commonalities between different kinds of data-analysis libraries, we found that there is a strong correlation between the number of bugs in data acquisition and visualization packages. 
Furthermore, the \igraph package in R is less correlated to other packages. For the mathematical function and machine learning packages, the equivalent packages are also strongly correlated (in~\figref{fig:coefVizMath}(b)) as they are providing similar functionality. The random forest has low correlations with other packages and libraries other than \rcaret. We further investigated and identified common antipatterns~\cite{antipatterns}, e.g., Cut-and-Paste Programming and Spaghetti Code, that explain these strong correlations among libraries.

\textbf{Implication:}
\textit{Researchers can further investigate the correlations between bugs and root causes among other R and Python packages. Similar bug detection and repair strategies could be leveraged, as bugs in R are likely to occur in Python or vice-versa.}

%% file: correlation.tex
\begin{figure}[h]
	\centering
	\subfloat[]{{\includegraphics[width=3.8cm, trim={0.2cm 2.2cm 0cm 3.8cm},clip]{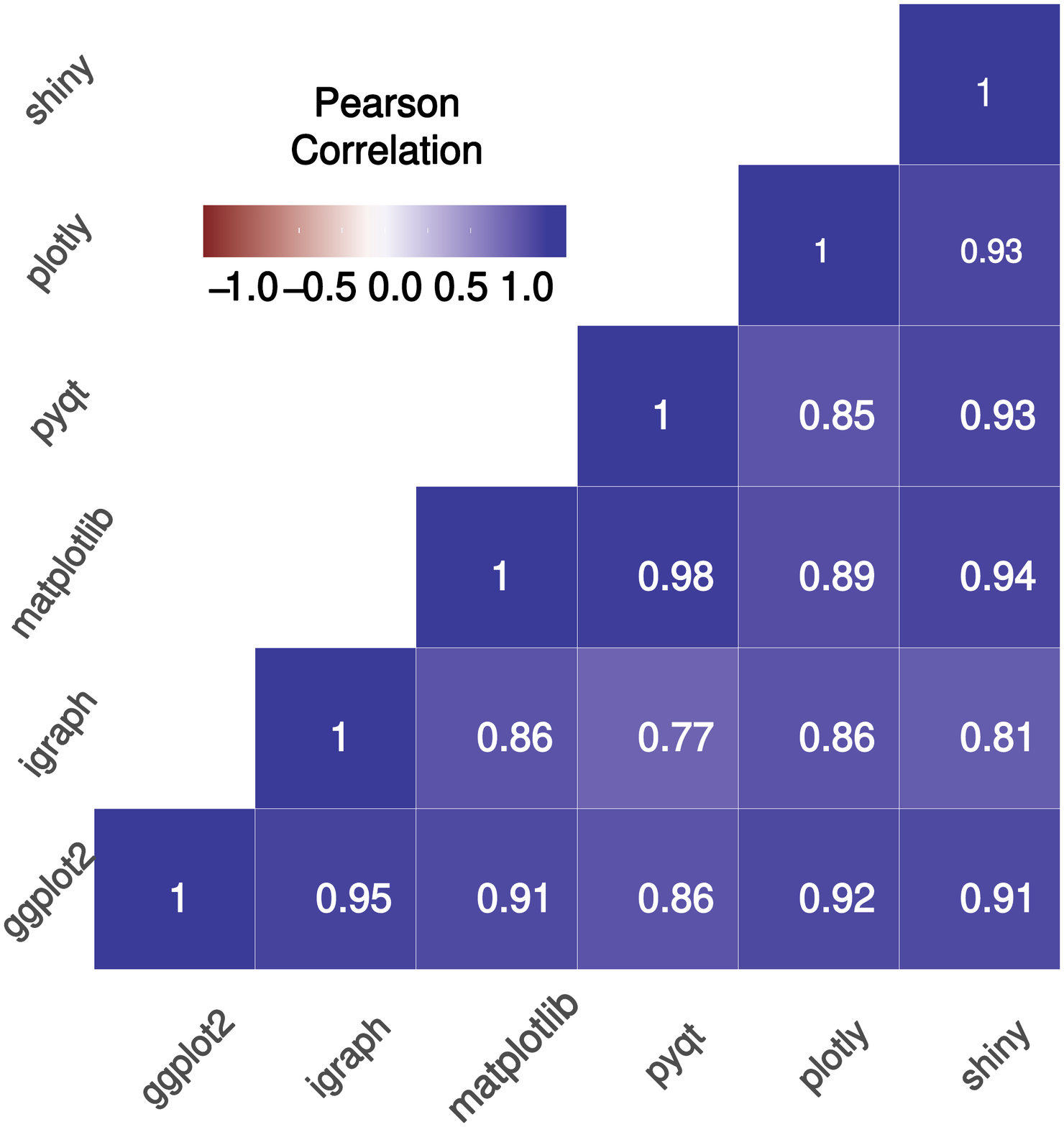} }}%
	\qquad
	\subfloat[]{{\includegraphics[width=3.8cm,trim={0cm 2.5cm 0cm 4cm},clip]{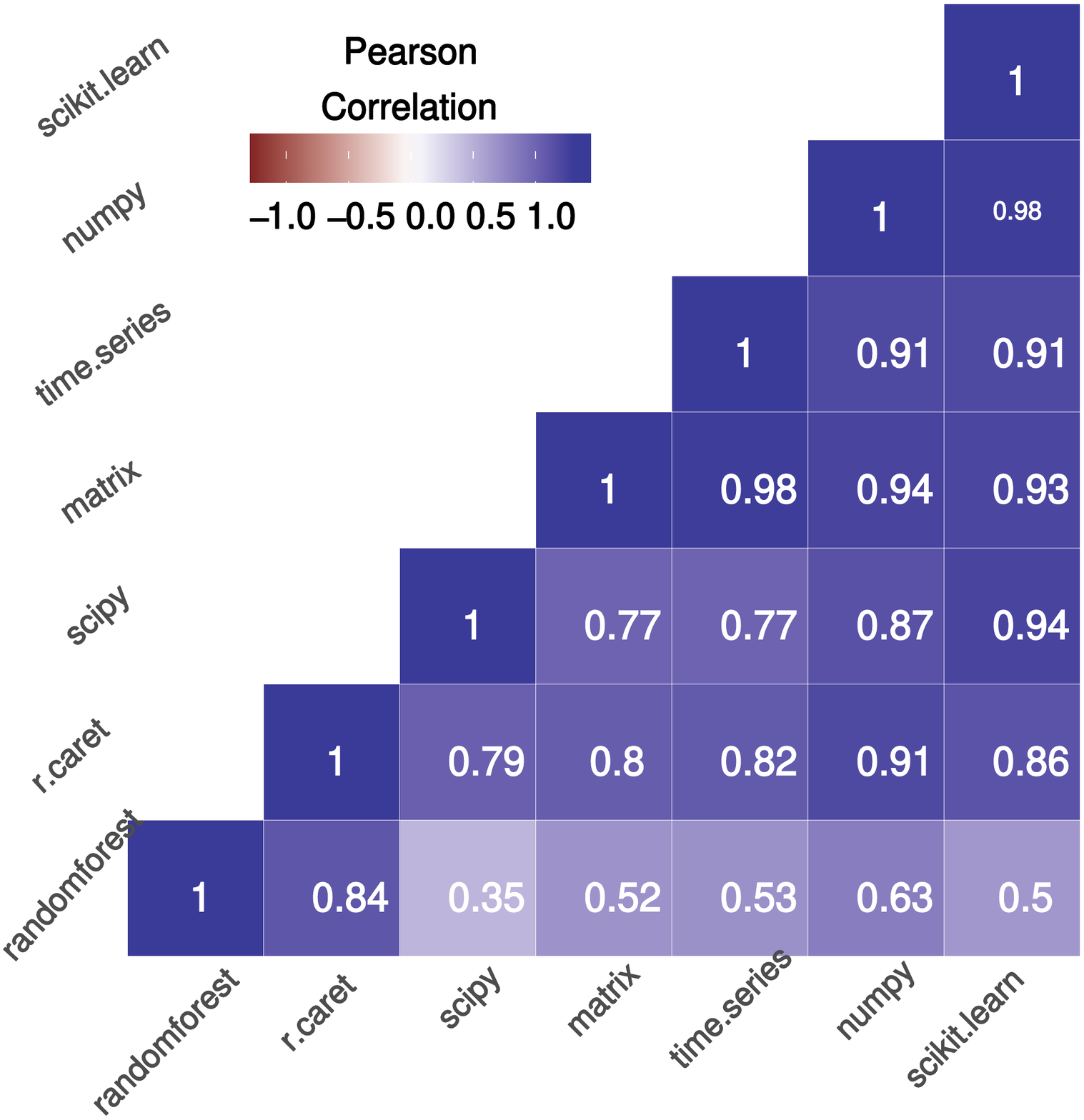} }}%
	\caption{The correlation of bug types among similar libraries in R and Python for (a) visualization, and (b) mathematical and machine learning packages}%
	\label{fig:coefVizMath}%
\end{figure}

%% file: lessons.tex
\section{Lessons Learned}
\label{sec:lessons}

Bugs in data analytics programs differ from traditional programs since they require addressing not only the code but also data quality, probabilistic and big data, statistical/mathematical, and domain knowledge. This study opens up several actionable directions for the less explored area, i.e., software engineering for data science. The key findings of RQ1 indicate that Python is a better language than R for data analytics w.r.t. having fewer bugs, especially, Data bugs, Data Flow Bugs, IIS Bugs, and Initialization Bugs.



\textbf{Suggestions for researchers:}
First, the dataset could be utilized to identify bug-fix patterns in data analytics programs and fed into automatic repair tools. Second, to help reduce confusion in data analytics tasks, package vendors should complement the documentation and specification by following our labeled set of bugs as a reference. Third, data analytics pipelines need abstract representation that benefits developers for dimension mismatch problems and improves the reusability of the data analytics pipeline. Fourth, the profiling tools also can benefit from performance-related bugs we found. Our dataset could be a reference point for different parameters and APIs choices that affected pipeline performance. Finally, we encouraged researchers to design and develop bug detection and localization approaches which can leverage crashes as an effect to detect a significant fraction of the bugs in data analytics programs.

\textbf{Suggestions for practitioners:}
First, developers of data analytics programs can learn from the common mistakes that others make and we call for data analytics educators to incorporate more SE concepts into their curriculum. Second, we also call the SE community for software flaws e.g., bad code smell and anti-patterns, the necessity of automatic testing tools, better deprecation strategies, declarative techniques, and automatic visualization for data analytics programs. Third, to mitigate bugs in data analytics packages, developers should be aware of incorrect data analysis parameters pointed out herein. Finally, researchers could determine if there is a correlation between bugs and root causes among R and Python data analytics packages. So that detection and repair strategies of those kinds of bugs in one language can be enriched by the other.


%% file: tv.tex
\section{Threats to Validity}
\label{sec:tv}

{\bf Internal threat.\ }
The results obtained in this paper could be biased by fundamental differences between R and Python. 
These differences in programming philosophy have the potential to affect the results. To mitigate this issue, we have chosen the most frequent packages in similar tasks in data analytics. There are also highly used packages in R that do not have a similar library in Python and vice versa. Bugs in the library are faults in the source code of the library while ``bugs in user code'' are faults that users make when using a specific library~\cite{jahangirova2019taxonomy}. Our work focuses on bugs in user code, bugs in libraries~\cite{compilerbug} can be investigated in future work.

{\bf External threat.\ }
The trustworthiness of the dataset we collected could be a threat to the validity of our results. To address this threat, we have chosen the high-score posts from \sof that answered by highly popular developers and also accepted by the person who asked the question. Some of these bug fixes may cause another bug that needs further investigation. The classifications were done by authors incorporating independent labeling after reconciling differences with an expert. To ensure that findings are statistically significant, we have conducted the Mann-Whitney U-test. We found similar result for \sof and \gh datasets, i.e., the distribution of both findings is statistically significant.

%% file: related.tex
\section{Related Works}
\textbf{Empirical Study on Bugs.} Programming bugs are a well-studied field in software engineering. There are multiple attempts to find empirical evidence of bugs in different programming languages and platforms e.g., Java ~\cite{saha2018bugs, ebert2015exploratory}, Python ~\cite{ma2017developers}, Firefox ~\cite{zaman2011security}, MySQL, Apache, Mozilla and OpenOffice ~\cite{lu2008learning}, Mozzila and Apache open source systems ~\cite{li2006have}, and operating system ~\cite{chou2001empirical}. There are empirical studies on bugs in machine learning and deep learning programs~\cite{islam2019comprehensive, zhang2018empirical, thung2012empirical,zhangempirical,islam20repairing,jahangirova2019taxonomy, garcia2020comprehensive} and data science pipelines~\cite{biswas2022art}. These studies have been conducted on real-life examples from the \sof post and \gh commits. Thung \etal studied ML bugs of three systems, Apache Mahout, Lucene, and OpenNLP~\cite{thung2012empirical}, in which they classified the bugs into six categories of bug frequencies, bug types, the severity of the bug, bug-fixing duration, bug-fixing effort, and impact of bugs. There are prior works~\cite{PySStuBs, BugsInPy} on characterizing bugs in popular open-source python projects to enable controlled testing and debugging studies. Our work focuses on characterizing and investigating 8 different bugs types in data analytics programs. Zhang \etal utilized \sof questions, and \gh commits to investigate bugs in deep learning applications built on top of \tensor ~\cite{zhang2018empirical}. They focused on symptoms and root causes of Tensorflow bugs to have a better understanding of deep learning bugs. The dataset and methodology of this paper are inspired by a deep learning bug study based on \sof questions, and mining Github commits ~\cite{islam2019comprehensive}. 
They also adapted a taxonomy of bug type, root cause, and bug impacts for deep learning software.
However, these bugs related studies are not focusing on data analytics programs. 
Claes \etal~\cite{claes2014maintainability} have conducted a study on over 5000 R packages and have explored the dependency error, fixes, and maintainability of packages in R. Decan \etal~\cite{decan2015development} have done an empirical study on R ecosystems, and they have analyzed the development and distribution of R packages on \CRAN , \BioConductor, \gh, and \RForge. Package dependency in three programming languages, i.e., R, Python, and JavaScript, are studied~\cite{decan2016topology}. 
However, to the best of our knowledge, there has not been a large scale bug study on data analytics programs. 

\textbf{Stack Overflow and Github study.} 
Kavaler~\etal used \sof posts to understand the difficulties faced by developers to use Android APIs ~\cite{kavaler2013using}. In another work, the 3,474,987 posts were studied to understand the trend in \sof and how this varies over the period ~\cite{barua2014developers}. Parnin~\etal analyzed \sof posts to investigate the coverage of the posts' discussions  ~\cite{parnin2012crowd}. Ma \etal~\cite{ma2017developers} studied \gh issues from 271 Python-based projects and reported frequently used practices carried out by developers from these projects in addressing issues. There are \sof posts' study~\cite{bagherzadeh2019going} for big data code and learning and debugging Rust safety rules~\cite{zhu2022learning}; however, there exists no large scale bug study on data analytics programs.

%% file: conclusion.tex
\section{Conclusion and Future Work}
\label{sec:conclusion}
In this work, we present an in-depth analysis of the bugs that programmers encounter when writing data analytics programs. This study investigates R and Python programs in three curated datasets, which are composed of 5,068 \sof posts, 1,800 \gh fix commits, and \gh issues of data analytics packages. The goal of the study is to understand the landscape and frequency of bugs in data analytics programs. Building on prior bug taxonomies, this work yielded categorizations of bug types, the root causes, and the effects that data analytics bug types can have on program execution. We also report which bug types and root causes are most frequent, the relationships of bug types and effects, and provide implications for the findings. Our study offers implications for researchers and practitioners in data analytics. For example, this study highlights the importance of abstract representations, profiling tools, and performance-related bug analysis for data analytics programs. Our dataset can help researchers analyze bug-fix patterns and enhance automatic repair tools. Furthermore, our dataset can be used as reference documentation and specification for data analytics programs. For practitioners, our study emphasizes the need for incorporating software engineering concepts into data analytics education and the use of automatic testing tools, improved deprecation strategies, and awareness of incorrect data analysis parameters. Finally, we found that despite the methodological and fundamental differences between R and Python, there is a strong correlation between comparable packages. Thus, repair strategies for R can be applied for Python and vice versa.

%% file: data-availability.tex
\section{Data Availability}
\label{sec:datavail}
The benchmark datasets and results are available in this anonymous GitHub repository~\cite{dataset} that offers valuable resources for future research.


%% file: EMSE23RPython.bbl
\begin{thebibliography}{97}
\providecommand{\natexlab}[1]{#1}
\providecommand{\url}[1]{{#1}}
\providecommand{\urlprefix}{URL }
\expandafter\ifx\csname urlstyle\endcsname\relax
  \providecommand{\doi}[1]{DOI~\discretionary{}{}{}#1}\else
  \providecommand{\doi}{DOI~\discretionary{}{}{}\begingroup
  \urlstyle{rm}\Url}\fi
\providecommand{\eprint}[2][]{\url{#2}}

\bibitem[{plm(2016)}]{plmc2}
 (2016) {Drop stringi dependency}.
  \url{https://github.com/tidyverse/tidyr/issues/936}, [Online; accessed
  May-2023]

\bibitem[{dfb(2016)}]{dfbissue2}
 (2016) {Git issues about dplyr debugger}.
  \url{https://github.com/tidyverse/dplyr/issues/2295}, [Online; accessed
  May-2023]

\bibitem[{dbi(2016)}]{dbissue2}
 (2016) {Import multiple CSVs without headers into single R data frame}.
  \url{https://stackoverflow.com/questions/40348902/}, [Online; accessed
  May-2023]

\bibitem[{plm(2017)}]{plmc5}
 (2017) {dplyr 0.5.0.9002 prevents CRAN version of sparklyr from loading}.
  \url{https://github.com/tidyverse/dplyr/issues/2670}, [Online; accessed
  May-2023]

\bibitem[{dbi(2018)}]{dbissue1}
 (2018) {Read.table with no header from Rstudio Community }.
  \url{https://community.rstudio.com/t/read-table-with-no-header/11459},
  [Online; accessed May-2023]

\bibitem[{so2(2019{\natexlab{a}})}]{so26244321}
 (2019{\natexlab{a}}) {dplyr summarise multiple columns using t.test}.
  \url{https://stackoverflow.com/questions/26244321/}, [Online; accessed
  May-2023]

\bibitem[{so3(2019{\natexlab{a}})}]{so30562819}
 (2019{\natexlab{a}}) {Error message when running simple 'rename' function in
  R}. \url{https://stackoverflow.com/questions/30562819/}, [Online; accessed
  May-2023]

\bibitem[{so3(2019{\natexlab{b}})}]{so31717850}
 (2019{\natexlab{b}}) {Error: package or namespace load failed for ggplot2 and
  for data.table}. \url{https://stackoverflow.com/questions/31717850/},
  [Online; accessed May-2023]

\bibitem[{so9(2019)}]{so9652832}
 (2019) {How to load a tsv file into a Pandas DataFrame?}
  \url{https://stackoverflow.com/questions/9652832/}, [Online; accessed
  May-2023]

\bibitem[{192(2019)}]{19202093}
 (2019) {How to replace NaN values by Zeroes in a column of a Pandas
  Dataframe?}
  \url{https://stackoverflow.com/questions/19202093/how-to-select-columns-from-groupby-object-in-pandas},
  [Online; accessed May-2023]

\bibitem[{so4(2019)}]{so48066517}
 (2019) {Python: Pandas pd.read_excel giving ImportError: Install xlrd $\ge$
  0.9.0 for Excel support}.
  \url{https://stackoverflow.com/questions/48066517/}, [Online; accessed
  May-2023]

\bibitem[{so2(2019{\natexlab{b}})}]{so26535563}
 (2019{\natexlab{b}}) {Querying for NaN and other names in Pandas}.
  \url{https://stackoverflow.com/questions/26535563/}, [Online; accessed
  May-2023]

\bibitem[{rDo(2019)}]{rDocumentation}
 (2019) {R Documentation }. \url{https://www.rdocumentation.org/}, [Online;
  accessed May-2023]

\bibitem[{261(2019)}]{26106146}
 (2019) {Why does summarize or mutate not work with group_by when I load `plyr`
  after `dplyr`?} \url{https://stackoverflow.com/questions/26106146}, [Online;
  accessed May-2023]

\bibitem[{bbc(2020)}]{bbcAI}
 (2020) {AAAS: Machine learning causing science crisis}.
  \url{https://www.bbc.com/news/science-environment-47267081}, [Online;
  accessed May-2023]

\bibitem[{khn(2020)}]{khnMedical}
 (2020) {Death By 1,000 Clicks: Where Electronic Health Records Went Wrong}.
  \url{https://khn.org/news/death-by-a-thousand-clicks/}, [Online; accessed
  May-2023]

\bibitem[{plm(2020{\natexlab{a}})}]{plmc4}
 (2020{\natexlab{a}}) {Drop purrr dependency}.
  \url{https://github.com/tidyverse/tidyr/issues/941}, [Online; accessed
  May-2023]

\bibitem[{git(2020{\natexlab{a}})}]{gitissueCSVr}
 (2020{\natexlab{a}}) {Git issues about readr of tidyverse}.
  \url{https://github.com/tidyverse/readr/issues}, [Online; accessed May-2023]

\bibitem[{dfb(2020)}]{dfbissue1}
 (2020) {Git issues demanding debugger watch window}.
  \url{https://github.com/rstudio/rstudio/issues/8555}, [Online; accessed
  May-2023]

\bibitem[{so2(2020)}]{so21737815}
 (2020) {Grouped operations that result in length not equal to 1 or length of
  group in dplyr}. \url{https://stackoverflow.com/questions/21737815/},
  [Online; accessed May-2023]

\bibitem[{dat(2020)}]{dataversity}
 (2020) {How is Bad Data Crippling Your Data Analytics?}
  \url{https://www.dataversity.net/bad-data-crippling-data-analytics/},
  [Online; accessed May-2023]

\bibitem[{so1(2020)}]{so15138973}
 (2020) {How to get the number of the most frequent value in a column?}
  \url{https://stackoverflow.com/questions/15138973}, [Online; accessed
  May-2023]

\bibitem[{nue(2020)}]{nuemdMedical}
 (2020) {Medical misdiagnosis is a serious issue, but EHR software and
  databases may have a solution}.
  \url{https://www.nuemd.com/news/2013/09/10/medical-misdiagnosis-is-a-serious-issue-but-ehr-software-and-databases-may-have-a-solution/},
  [Online; accessed May-2023]

\bibitem[{git(2020{\natexlab{b}})}]{gitissueCSVnrows}
 (2020{\natexlab{b}}) {read_csv with filehandler and nrows argument}.
  \url{https://github.com/pandas-dev/pandas/issues/17155}, [Online; accessed
  May-2023]

\bibitem[{plm(2020{\natexlab{b}})}]{plmc1}
 (2020{\natexlab{b}}) {Several functions won't work if you don't load dplyr}.
  \url{https://github.com/tidyverse/dplyr/issues/2297}, [Online; accessed
  May-2023]

\bibitem[{yah(2020)}]{yahooFairness}
 (2020) {The Great Algorithm Problem: How Do We Know They Will Be Fair?}
  \url{https://news.yahoo.com/great-algorithm-problem-know-fair-081100779.html},
  [Online; accessed May-2023]

\bibitem[{dfb(2021)}]{dfbissue3}
 (2021) {Git issues about non-joined duplicate variables}.
  \url{https://github.com/tidyverse/dplyr/issues/5860}, [Online; accessed
  May-2023]

\bibitem[{plm(2022)}]{plmc3}
 (2022) {Memory leak in dplyr, magrittr, or one of dplyr's dependencies}.
  \url{https://github.com/tidyverse/dplyr/issues/6207}, [Online; accessed
  May-2023]

\bibitem[{dat(2023)}]{dataset}
 (2023) {Dataset of Data Analytics Bugs in R and Python}.
  \url{https://github.com/DataAnalyticsProgram/RPython}, [Online; accessed
  May-2023]

\bibitem[{{Alexander Shvets}(2017)}]{antipatterns}
{Alexander Shvets} (2017) {Software Development AntiPatterns}.
  \url{https://sourcemaking.com/antipatterns/software-development-antipatterns}

\bibitem[{Bagherzadeh and Khatchadourian(2019)}]{bagherzadeh2019going}
Bagherzadeh M, Khatchadourian R (2019) Going big: a large-scale study on what
  big data developers ask. In: Proceedings of the 2019 27th ACM Joint Meeting
  on European Software Engineering Conference and Symposium on the Foundations
  of Software Engineering, ACM, pp 432--442

\bibitem[{Barua et~al.(2014)Barua, Thomas, and Hassan}]{barua2014developers}
Barua A, Thomas SW, Hassan AE (2014) What are developers talking about? an
  analysis of topics and trends in stack overflow. Empirical Software
  Engineering 19(3):619--654

\bibitem[{Becker et~al.(1988)Becker, Chambers, and Wilks}]{data.table}
Becker RA, Chambers JM, Wilks AR (1988) The new s language. wadsworth \&
  brooks. Cole publication

\bibitem[{Beizer(1984)}]{beizer1984software}
Beizer B (1984) Software system testing and quality assurance. Van Nostrand
  Reinhold Co.

\bibitem[{Biswas et~al.(2022)Biswas, Wardat, and Rajan}]{biswas2022art}
Biswas S, Wardat M, Rajan H (2022) The art and practice of data science
  pipelines: A comprehensive study of data science pipelines in theory,
  in-the-small, and in-the-large. In: Proceedings of the 44th International
  Conference on Software Engineering, pp 2091--2103

\bibitem[{Chou et~al.(2001)Chou, Yang, Chelf, Hallem, and
  Engler}]{chou2001empirical}
Chou A, Yang J, Chelf B, Hallem S, Engler D (2001) An empirical study of
  operating systems errors. In: ACM SIGOPS Operating Systems Review, ACM,
  vol~35, pp 73--88

\bibitem[{Claes et~al.(2014)Claes, Mens, and
  Grosjean}]{claes2014maintainability}
Claes M, Mens T, Grosjean P (2014) On the maintainability of cran packages. In:
  2014 Software Evolution Week-IEEE Conference on Software Maintenance,
  Reengineering, and Reverse Engineering (CSMR-WCRE), IEEE, pp 308--312

\bibitem[{Csardi et~al.(2006)Csardi, Nepusz et~al.}]{igraph}
Csardi G, Nepusz T, et~al. (2006) The igraph software package for complex
  network research. InterJournal, Complex Systems 1695(5):1--9

\bibitem[{Decan et~al.(2015)Decan, Mens, Claes, and
  Grosjean}]{decan2015development}
Decan A, Mens T, Claes M, Grosjean P (2015) On the development and distribution
  of r packages: An empirical analysis of the r ecosystem. In: Proceedings of
  the 2015 european conference on software architecture workshops, ACM, p~41

\bibitem[{Decan et~al.(2016)Decan, Mens, and Claes}]{decan2016topology}
Decan A, Mens T, Claes M (2016) On the topology of package dependency networks:
  A comparison of three programming language ecosystems. In: Proccedings of the
  10th European Conference on Software Architecture Workshops, ACM, p~21

\bibitem[{Ebert et~al.(2015)Ebert, Castor, and
  Serebrenik}]{ebert2015exploratory}
Ebert F, Castor F, Serebrenik A (2015) An exploratory study on exception
  handling bugs in {Java} programs. Journal of Systems and Software 106:82--101

\bibitem[{Fincher and Tenenberg(2005)}]{fincher2005making}
Fincher S, Tenenberg J (2005) Making sense of card sorting data. Expert Systems
  22(3):89--93

\bibitem[{Fowler(2018)}]{fowler2018refactoring}
Fowler M (2018) Refactoring: improving the design of existing code.
  Addison-Wesley Professional

\bibitem[{Gao et~al.(2015)Gao, Zhang, Wang, Xiong, Zhang, and
  Mei}]{gao2015fixing}
Gao Q, Zhang H, Wang J, Xiong Y, Zhang L, Mei H (2015) {Fixing recurring crash
  bugs via analyzing Q\&A sites (T)}. In: 2015 30th IEEE/ACM International
  Conference on Automated Software Engineering (ASE), IEEE, pp 307--318

\bibitem[{Garcia et~al.(2020)Garcia, Feng, Shen, Almanee, Xia, and
  Chen}]{garcia2020comprehensive}
Garcia J, Feng Y, Shen J, Almanee S, Xia Y, Chen QA (2020) A comprehensive
  study of autonomous vehicle bugs. In: 2020 IEEE/ACM 42nd International
  Conference on Software Engineering (ICSE)

\bibitem[{Graham et~al.(2010)Graham, Furr, Kuczmarski, Biskup, and
  Palay}]{pycontracturl}
Graham B, Furr W, Kuczmarski K, Biskup B, Palay A (2010) Pycontracts.
  \urlprefix\url{https://andreacensi.github.io/contracts//}

\bibitem[{Herzig et~al.(2013)Herzig, Just, and
  Zeller}]{10.5555/2486788.2486840}
Herzig K, Just S, Zeller A (2013) It's not a bug, it's a feature: How
  misclassification impacts bug prediction. In: Proceedings of the 2013
  International Conference on Software Engineering, IEEE Press, ICSE ’13, p
  392–401

\bibitem[{Hong et~al.(2016)Hong, Ghosh, and Srinivasan}]{hong2016dealing}
Hong C, Ghosh R, Srinivasan S (2016) Dealing with class imbalance using
  thresholding. \eprint{1607.02705}

\bibitem[{Horton and Parnin(2019)}]{horton2019dockerizeme}
Horton E, Parnin C (2019) Dockerizeme: Automatic inference of environment
  dependencies for python code snippets. In: 2019 IEEE/ACM 41st International
  Conference on Software Engineering (ICSE), IEEE, pp 328--338

\bibitem[{Humbatova et~al.(2020)Humbatova, Jahangirova, , Bavota, Riccio,
  Stocco, and Tonella}]{jahangirova2019taxonomy}
Humbatova N, Jahangirova G, , Bavota G, Riccio V, Stocco A, Tonella P (2020)
  Taxonomy of real faults in deep learning systems. ICSE'20: The 42nd
  International Conference on Software Engineering

\bibitem[{Hunter(2007)}]{matplotlib}
Hunter JD (2007) Matplotlib: A 2d graphics environment. Computing in science \&
  engineering 9(3):90

\bibitem[{Ihaka and Gentleman(1996)}]{R}
Ihaka R, Gentleman R (1996) {R: A Language for Data Analysis and Graphics}.
  Journal of Computational and Graphical Statistics 5(3):299--314,
  \doi{10.1080/10618600.1996.10474713},
  \urlprefix\url{https://amstat.tandfonline.com/doi/abs/10.1080/10618600.1996.10474713},
  \eprint{https://amstat.tandfonline.com/doi/pdf/10.1080/10618600.1996.10474713}

\bibitem[{Islam et~al.(2019)Islam, Nguyen, Pan, and
  Rajan}]{islam2019comprehensive}
Islam MJ, Nguyen G, Pan R, Rajan H (2019) A comprehensive study on deep
  learning bug characteristics. In: Proceedings of the 2019 27th ACM Joint
  Meeting on European Software Engineering Conference and Symposium on the
  Foundations of Software Engineering, ACM, New York, NY, USA, ESEC/FSE 2019,
  pp 510--520, \doi{10.1145/3338906.3338955},
  \urlprefix\url{http://doi.acm.org/10.1145/3338906.3338955}

\bibitem[{Islam et~al.(2020)Islam, Pan, Nguyen, and Rajan}]{islam20repairing}
Islam MJ, Pan R, Nguyen G, Rajan H (2020) Repairing deep neural networks: Fix
  patterns and challenges. In: ICSE'20: The 42nd International Conference on
  Software Engineering

\bibitem[{Jones et~al.(2001)Jones, Oliphant, and Peterson}]{scipy}
Jones E, Oliphant T, Peterson P (2001) Scipy: Open source scientific tools for
  {Python}

\bibitem[{Kamienski et~al.(2021)Kamienski, Palechor, Bezemer, and
  Hindle}]{PySStuBs}
Kamienski AV, Palechor L, Bezemer CP, Hindle A (2021) Pysstubs: Characterizing
  single-statement bugs in popular open-source python projects. In: 2021
  IEEE/ACM 18th International Conference on Mining Software Repositories (MSR),
  pp 520--524, \doi{10.1109/MSR52588.2021.00066}

\bibitem[{Kavaler et~al.(2013)Kavaler, Posnett, Gibler, Chen, Devanbu, and
  Filkov}]{kavaler2013using}
Kavaler D, Posnett D, Gibler C, Chen H, Devanbu P, Filkov V (2013) Using and
  asking: {APIs} used in the android market and asked about in stackoverflow.
  In: International Conference on Social Informatics, Springer, pp 405--418

\bibitem[{{Kim} et~al.(2018){Kim}, {Zimmermann}, {DeLine}, and
  {Begel}}]{8046093}
{Kim} M, {Zimmermann} T, {DeLine} R, {Begel} A (2018) Data scientists in
  software teams: State of the art and challenges. IEEE Transactions on
  Software Engineering 44(11):1024--1038, \doi{10.1109/TSE.2017.2754374}

\bibitem[{Kuhn et~al.(2008)}]{caret}
Kuhn M, et~al. (2008) Building predictive models in r using the caret package.
  Journal of statistical software 28(5):1--26

\bibitem[{Li et~al.(2006)Li, Tan, Wang, Lu, Zhou, and Zhai}]{li2006have}
Li Z, Tan L, Wang X, Lu S, Zhou Y, Zhai C (2006) Have things changed now?: an
  empirical study of bug characteristics in modern open source software. In:
  Proceedings of the 1st workshop on Architectural and system support for
  improving software dependability, ACM, pp 25--33

\bibitem[{Liaw et~al.(2002)Liaw, Wiener et~al.}]{randomForest}
Liaw A, Wiener M, et~al. (2002) Classification and regression by randomforest.
  R news 2(3):18--22

\bibitem[{Lu et~al.(2008)Lu, Park, Seo, and Zhou}]{lu2008learning}
Lu S, Park S, Seo E, Zhou Y (2008) Learning from mistakes: a comprehensive
  study on real world concurrency bug characteristics. In: ACM SIGARCH Computer
  Architecture News, ACM, vol~36, pp 329--339

\bibitem[{Ma et~al.(2017)Ma, Chen, Zhang, Zhou, and Xu}]{ma2017developers}
Ma W, Chen L, Zhang X, Zhou Y, Xu B (2017) How do developers fix cross-project
  correlated bugs? a case study on the github scientific {Python} ecosystem.
  In: 2017 IEEE/ACM 39th International Conference on Software Engineering
  (ICSE), IEEE, pp 381--392

\bibitem[{McKinney(2011)}]{pandas}
McKinney W (2011) pandas: a foundational python library for data analysis and
  statistics. {Python} for High Performance and Scientific Computing 14

\bibitem[{McKinney et~al.(2010)}]{dataframe}
McKinney W, et~al. (2010) Data structures for statistical computing in
  {Python}. In: Proceedings of the 9th Python in Science Conference, Austin,
  TX, vol 445, pp 51--56

\bibitem[{Meyer(1992)}]{meyer1992applying}
Meyer B (1992) Applying'design by contract'. Computer 25(10):40--51

\bibitem[{Mitchell(1999)}]{mitchell1999machine}
Mitchell TM (1999) Machine learning and data mining. Communications of the ACM
  42(11):30--36

\bibitem[{{Nguyen} et~al.(2019){Nguyen}, {Di Rocco}, {Di Ruscio}, {Ochoa},
  {Degueule}, and {Di Penta}}]{focus}
{Nguyen} PT, {Di Rocco} J, {Di Ruscio} D, {Ochoa} L, {Degueule} T, {Di Penta} M
  (2019) Focus: A recommender system for mining api function calls and usage
  patterns. In: 2019 IEEE/ACM 41st International Conference on Software
  Engineering (ICSE), pp 1050--1060

\bibitem[{Parnin et~al.(2012)Parnin, Treude, Grammel, and
  Storey}]{parnin2012crowd}
Parnin C, Treude C, Grammel L, Storey MA (2012) Crowd documentation: Exploring
  the coverage and the dynamics of api discussions on stack overflow. Georgia
  Institute of Technology, Tech Rep 11

\bibitem[{Pedregosa et~al.(2011)Pedregosa, Varoquaux, Gramfort, Michel,
  Thirion, Grisel, Blondel, Prettenhofer, Weiss, Dubourg et~al.}]{scikit}
Pedregosa F, Varoquaux G, Gramfort A, Michel V, Thirion B, Grisel O, Blondel M,
  Prettenhofer P, Weiss R, Dubourg V, et~al. (2011) Scikit-learn: Machine
  learning in {Python}. Journal of machine learning research 12(Oct):2825--2830

\bibitem[{Peruma et~al.(2020)Peruma, Almalki, Newman, Mkaouer, Ouni, and
  Palomba}]{tsDetect}
Peruma A, Almalki K, Newman CD, Mkaouer MW, Ouni A, Palomba F (2020) Tsdetect:
  An open source test smells detection tool. In: Proceedings of the 28th ACM
  Joint Meeting on European Software Engineering Conference and Symposium on
  the Foundations of Software Engineering, Association for Computing Machinery,
  New York, NY, USA, ESEC/FSE 2020, p 1650–1654,
  \doi{10.1145/3368089.3417921},
  \urlprefix\url{https://doi.org/10.1145/3368089.3417921}

\bibitem[{{R Core Team}(2019)}]{shiny}
{R Core Team} (2019) R: A Language and Environment for Statistical Computing. R
  Foundation for Statistical Computing, Vienna, Austria,
  \urlprefix\url{https://www.R-project.org/}

\bibitem[{Runkler(2020)}]{runkler2020data}
Runkler TA (2020) Data analytics. Springer

\bibitem[{Saha et~al.(2018)Saha, Lyu, Lam, Yoshida, and Prasad}]{saha2018bugs}
Saha R, Lyu Y, Lam W, Yoshida H, Prasad M (2018) Bugs. jar: A large-scale,
  diverse dataset of real-world {Java} bugs. In: 2018 IEEE/ACM 15th
  International Conference on Mining Software Repositories (MSR), IEEE, pp
  10--13

\bibitem[{Sanner et~al.(1999)}]{sanner1999python}
Sanner MF, et~al. (1999) {Python}: a programming language for software
  integration and development. J Mol Graph Model 17(1):57--61

\bibitem[{Sarsa et~al.(2022)Sarsa, Denny, Hellas, and Leinonen}]{codex}
Sarsa S, Denny P, Hellas A, Leinonen J (2022) Automatic generation of
  programming exercises and code explanations using large language models. In:
  Proceedings of the 2022 ACM Conference on International Computing Education
  Research - Volume 1, Association for Computing Machinery, New York, NY, USA,
  ICER '22, p 27–43, \doi{10.1145/3501385.3543957},
  \urlprefix\url{https://doi.org/10.1145/3501385.3543957}

\bibitem[{Shen et~al.(2021)Shen, Ma, Chen, Tian, Cheung, and
  Chen}]{compilerbug}
Shen Q, Ma H, Chen J, Tian Y, Cheung SC, Chen X (2021) A comprehensive study of
  deep learning compiler bugs. In: Proceedings of the 29th ACM Joint Meeting on
  European Software Engineering Conference and Symposium on the Foundations of
  Software Engineering, Association for Computing Machinery, New York, NY, USA,
  ESEC/FSE 2021, p 968–980, \doi{10.1145/3468264.3468591},
  \urlprefix\url{https://doi.org/10.1145/3468264.3468591}

\bibitem[{Sievert(2018)}]{plotly}
Sievert C (2018) plotly for R. \urlprefix\url{https://plotly-r.com}

\bibitem[{Summerfield(2007)}]{pyqt}
Summerfield M (2007) Rapid GUI programming with {Python} and {Qt}: the
  definitive guide to PyQt programming. Pearson Education

\bibitem[{Thung et~al.(2012)Thung, Wang, Lo, and Jiang}]{thung2012empirical}
Thung F, Wang S, Lo D, Jiang L (2012) An empirical study of bugs in machine
  learning systems. In: 2012 IEEE 23rd International Symposium on Software
  Reliability Engineering, IEEE, pp 271--280

\bibitem[{Ushey et~al.(2015)Ushey, McPherson, Cheng, Atkins, and
  Allaire}]{ushey2015packrat}
Ushey K, McPherson J, Cheng J, Atkins A, Allaire J (2015) packrat: A dependency
  management system for projects and their {R} package dependencies. R package
  version 3

\bibitem[{Van Der~Walt et~al.(2011)Van Der~Walt, Colbert, and
  Varoquaux}]{numpy}
Van Der~Walt S, Colbert SC, Varoquaux G (2011) The numpy array: a structure for
  efficient numerical computation. Computing in Science \& Engineering 13(2):22

\bibitem[{Viera et~al.(2005)Viera, Garrett et~al.}]{viera2005understanding}
Viera AJ, Garrett JM, et~al. (2005) Understanding interobserver agreement: the
  kappa statistic. Fam med 37(5):360--363

\bibitem[{Wang et~al.(2020{\natexlab{a}})Wang, Li, and Zeller}]{jupyterStatic}
Wang J, Li L, Zeller A (2020{\natexlab{a}}) Better code, better sharing: On the
  need of analyzing jupyter notebooks. In: Proceedings of the ACM/IEEE 42nd
  International Conference on Software Engineering: New Ideas and Emerging
  Results, Association for Computing Machinery, New York, NY, USA, ICSE-NIER
  '20, p 53–56, \doi{10.1145/3377816.3381724},
  \urlprefix\url{https://doi.org/10.1145/3377816.3381724}

\bibitem[{Wang et~al.(2018)Wang, Wen, Liu, Wu, Wang, Yang, Yu, Zhu, and
  Cheung}]{wang2018dependency}
Wang Y, Wen M, Liu Z, Wu R, Wang R, Yang B, Yu H, Zhu Z, Cheung SC (2018) Do
  the dependency conflicts in my project matter? In: Proceedings of the 2018
  26th ACM Joint Meeting on European Software Engineering Conference and
  Symposium on the Foundations of Software Engineering, ACM, pp 319--330

\bibitem[{Wang et~al.(2020{\natexlab{b}})Wang, Wen, Liu, Wang, Li, Wang, Yu,
  Cheung, Xu, and Zhu}]{wang2020watchman}
Wang Y, Wen M, Liu Y, Wang Y, Li Z, Wang C, Yu H, Cheung SC, Xu C, Zhu Z
  (2020{\natexlab{b}}) Watchman: Monitoring dependency conflicts for python
  library ecosystem. In: Proceedings of the ACM/IEEE 42nd International
  Conference on Software Engineering, pp 125--135

\bibitem[{Warner and Guo(2017)}]{codepilot}
Warner J, Guo PJ (2017) Codepilot: Scaffolding end-to-end collaborative
  software development for novice programmers. In: Proceedings of the 2017 CHI
  Conference on Human Factors in Computing Systems, pp 1136--1141

\bibitem[{Wickham(2016)}]{ggplot2}
Wickham H (2016) ggplot2: elegant graphics for data analysis. Springer

\bibitem[{Wickham and Henry(2017)}]{tidyr}
Wickham H, Henry L (2017) Tidyr: Easily tidy data with spread () and gather ()
  functions. R package version 06 1

\bibitem[{Wickham et~al.(2023)}]{dplyr}
Wickham H, et~al. (2023) dplyr: A grammar of data manipulation

\bibitem[{Widyasari et~al.(2020)Widyasari, Sim, Lok, Qi, Phan, Tay, Tan, Wee,
  Tan, Yieh, Goh, Thung, Kang, Hoang, Lo, and Ouh}]{BugsInPy}
Widyasari R, Sim SQ, Lok C, Qi H, Phan J, Tay Q, Tan C, Wee F, Tan JE, Yieh Y,
  Goh B, Thung F, Kang HJ, Hoang T, Lo D, Ouh EL (2020) Bugsinpy: A database of
  existing bugs in python programs to enable controlled testing and debugging
  studies. In: Proceedings of the 28th ACM Joint Meeting on European Software
  Engineering Conference and Symposium on the Foundations of Software
  Engineering, Association for Computing Machinery, New York, NY, USA, ESEC/FSE
  2020, p 1556–1560, \doi{10.1145/3368089.3417943},
  \urlprefix\url{https://doi.org/10.1145/3368089.3417943}

\bibitem[{Yang et~al.(2022)Yang, Brower-Sinning, Lewis, and
  K{\"a}stner}]{dataleakage}
Yang C, Brower-Sinning RA, Lewis GA, K{\"a}stner C (2022) Data leakage in
  notebooks: Static detection and better processes. Proceedings of the 37th
  ACM/IEEE International Conference on Automated Software Engineering

\bibitem[{Zaman et~al.(2011)Zaman, Adams, and Hassan}]{zaman2011security}
Zaman S, Adams B, Hassan AE (2011) Security versus performance bugs: a case
  study on firefox. In: Proceedings of the 8th working conference on mining
  software repositories, ACM, pp 93--102

\bibitem[{Zhang et~al.(2012)Zhang, Yang, Zhang, Fan, Zhang, Zhao, and
  Ou}]{zhang2012automatic}
Zhang C, Yang J, Zhang Y, Fan J, Zhang X, Zhao J, Ou P (2012) Automatic
  parameter recommendation for practical api usage. In: Proceedings of the 34th
  International Conference on Software Engineering, IEEE Press, pp 826--836

\bibitem[{Zhang et~al.(2020)Zhang, Xiao, Zhang, Liu, Lin, and
  Yang}]{zhangempirical}
Zhang R, Xiao W, Zhang H, Liu Y, Lin H, Yang M (2020) An empirical study on
  program failures of deep learning jobs. ICSE'20: The 42nd International
  Conference on Software Engineering

\bibitem[{Zhang et~al.(2018)Zhang, Chen, Cheung, Xiong, and
  Zhang}]{zhang2018empirical}
Zhang Y, Chen Y, Cheung SC, Xiong Y, Zhang L (2018) An empirical study on
  tensorflow program bugs. In: Proceedings of the 27th ACM SIGSOFT
  International Symposium on Software Testing and Analysis, ACM, pp 129--140

\bibitem[{Zhu et~al.(2022)Zhu, Zhang, Qin, Xiong, and Song}]{zhu2022learning}
Zhu S, Zhang Z, Qin B, Xiong A, Song L (2022) Learning and programming
  challenges of rust: A mixed-methods study. In: Proceedings of the 44th
  International Conference on Software Engineering, pp 1269--1281

\end{thebibliography}
